\documentclass[acmtog,screen,nonacm]{acmart}
\AtBeginDocument{%
  \providecommand\BibTeX{{%
    \normalfont B\kern-0.5em{\scshape i\kern-0.25em b}\kern-0.8em\TeX}}}

\setcopyright{none}

\newcommand{\sys}{\mbox{LIMEADE}}
\newcommand{\sanity}{Semantic Sanity}
\newcommand{\etal}{\textit{et~al.}}
\newcommand{\eg}{\textit{e.g.}}

\usepackage{xcolor}
\usepackage{algorithm}
\usepackage{algorithmic}
\usepackage{subcaption}
\usepackage{booktabs}
\usepackage{amsfonts}

\begin{document}

\title{LIMEADE: From AI Explanations to Advice Taking}

\author{Benjamin Charles Germain Lee}
\authornote{Work performed during internship at the Allen Institute for Artificial Intelligence and Ph.D. at the University of Washington.}
\email{bcgl@cs.washington.edu}
\affiliation{%
  \institution{University of Washington \& Allen Institute for Artificial Intelligence}
  \country{USA}
}

\author{Doug Downey}
\email{dougd@allenai.org}
\affiliation{%
  \institution{Northwestern University \& Allen Institute for Artificial Intelligence}
  \country{USA}
}

\author{Kyle Lo}
\email{kylel@allenai.org}
\affiliation{%
  \institution{Allen Institute for Artificial Intelligence}
  \country{USA}
}

\author{Daniel S. Weld}
\email{weld@cs.washington.edu}
\affiliation{%
  \institution{University of Washington \& Allen Institute for Artificial Intelligence}
  \country{USA}
}

\renewcommand{\shortauthors}{Lee, Downey, Lo, and Weld}

\begin{abstract}
Research in human-centered AI has shown the benefits of systems that can explain their predictions. Methods that allow an AI to take advice from humans in response to explanations are similarly useful. While both capabilities are well-developed for \textit{transparent} learning models (e.g., linear models and GA$^2$Ms), and recent techniques (e.g., LIME and SHAP) can generate explanations for \textit{opaque} models, little attention has been given to advice methods for opaque models. This paper introduces \sys, the first general framework that translates  both positive and negative advice (expressed using  high-level vocabulary such as that employed by post-hoc explanations) into an update to an arbitrary, underlying opaque model. We demonstrate the generality of our approach with case studies on seventy real-world models across two broad domains: image classification and text recommendation. We show our method improves accuracy compared to a rigorous baseline on the image classification domains. For the text modality, we apply our framework to a neural recommender system for scientific papers on a public website;  our user study shows that our framework leads to significantly higher perceived user control, trust, and satisfaction. 
\end{abstract}



\maketitle

\section{Introduction}

A long-standing vision in AI is the construction of an {\em advice taker}, a system whose behavior, in the words of John McCarthy, ``will be improvable merely by making statements to it, telling it about its symbolic environment and what is wanted from it. To make these statements will require little if any knowledge of the program or the previous knowledge of the advice taker''~\cite{McCarthy1960ProgramsWC}. Indeed, today's guidelines for human-AI interaction dictate that ML systems should be able to explain their predictions to end-users and accept advice and corrections from them ~\cite{Amershi2019GuidelinesFH, amershi_power_2014, rao_2022}. Both explanation and advice-taking methods exist for transparent models, such as linear classifiers or generalized additive models (G$\text{A}^2$Ms) \cite{caruana_3, wang2021, sherry_tochi}, and their benefits for transparent recommenders have been demonstrated within the human-in-the-loop machine learning and human-AI interaction literature \cite{bostandjiev_tasteweights_2012, kulesza_tell_2012}.  These advice-taking approaches allow the human to provide high-level feedback on how specific input features should be driving the transparent model's behavior. In our related work (Section \ref{sec:relatedwork}), we elaborate on such approaches.

However, opaque models, such as boosted decision forests and deep neural networks, are a different story. Because they often provide the highest performance and are widely used, numerous researchers have investigated methods for generating post-hoc explanations of opaque ML models --- typically by creating a transparent approximation to the opaque model, called an explanatory model~\cite{survey_explaining_blackbox}. Several researchers have developed methods for translating high-level {\em  human advice} into specific classes of differentiable, neural models~\cite{doshivelez,liu,rieger,schramowski, doshi-velez_towards_2017}. However, to our knowledge only Schramowski \etal~\cite{schramowski} have introduced a method that works for {\em arbitrary} opaque models, and it is not capable of handling advice that corrects an agent's erroneous predictions (Section~\ref{sec:opaque-advice}).

Furthermore, even the advice-taking methods whose application is restricted to specific opaque model classes~\cite{liu,rieger,doshivelez} have limited empirical evaluation, often restricted to datasets that have been artificially biased (\eg, Decoy MNIST and Iris-Cancer~\cite{doshivelez}) in a way that a simple human tip (\eg, ``Ignore the artifact in the lower right corner'') can correct the problem. To demonstrate that advice-taking methods are useful in actual practice, experiments with large, real-world domains seem essential.   

Thus, two central questions for human-AI interaction remain unanswered:
\begin{enumerate}
    \item  Can one translate high-level human advice into a correction to an arbitrary, opaque, machine-learned model which uses a different set of features than those used to express the advice?
    \item Do these methods allow end-users to improve the accuracy of natural, real-world models more easily than by simply annotating more instances?
\end{enumerate}

\begin{figure*}
  \centering
\includegraphics[width=0.8\linewidth]{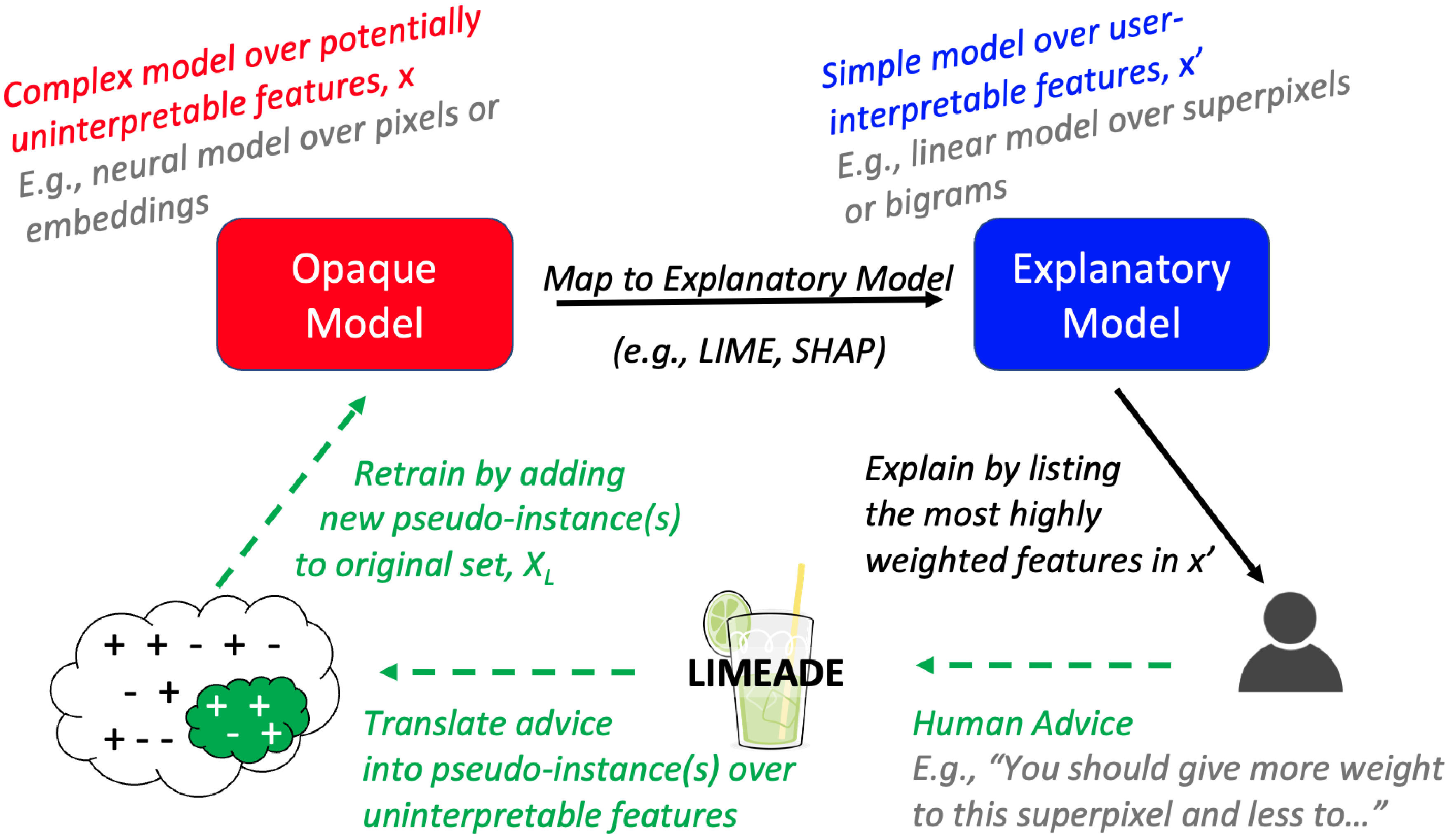}
\vspace*{-.2in}
  \caption{\sys\ takes a user's advice -- given in terms of features of the explanatory model --- and then modifies the original, opaque model by retraining. This is challenging because the mapping from opaque to explanatory model is typically many-to-one and hence not invertable.}
  \label{fig:framework}
\end{figure*}

This paper answers the first question affirmatively, but presents mixed results on the second.
Specifically, we present \sys, a general framework for updating an arbitrary, opaque machine learned model given high-level human advice, e.g. phrased in the same vocabulary used by a posthoc explanation of its behavior.  As shown in Figure \ref{fig:framework}, our approach builds upon explanatory approaches such as LIME \cite{Ribeiro_LIME_2016} and SHAP \cite{lundberg_2017} that describe the local behavior of a model in the region of a given instance. Given a trained model and an instance to be classified, these post-hoc approaches output an explanation in the form of a weighted list of {\em interpretable} features (typically distinct from the features utilized in the opaque model) that influence the instance's classification. With \sys, a user can then provide feedback in the same high-level terms as the explanation in order to modify the original, opaque model. \sys\ converts this user advice back into the original feature space of the opaque classifier by generating pseudo-instances representative of these features and retraining.  Unlike other methods intended for machine learning practitioners and model developers, \sys\ empowers end-users with little or no machine learning expertise to tune the system. 

\sys\ builds on the longstanding research areas of human-in-the-loop machine learning and interactive machine learning to provide a framework that is sufficiently general to address a wide range of model architectures, tasks, and modes of advice. We emphasize that \sys\ is  a general framework in three distinct senses: 
\begin{enumerate}
    \item \sys\ can be utilized for a wide range of advice-taking applications, from explanatory debugging to personalized recommendation.
    \item \sys\ is architecture-agnostic and enables advice taking for different types of opaque machine learning models, including both classifiers and rankers.
    \item \sys\ accepts different types of human advice (in this paper, we focus on advice given as binary feedback in terms of high-level features).
\end{enumerate}

Accordingly, we show that our framework is general by demonstrating its success on seventy real-world models across two broad domains: image classification and text ranking.
For our first case study, we use \sys\ to give advice to  twenty binary image classifiers (\eg, models predicting ``giraffe'' or ``not giraffe'') that are built on  precomputed neural embeddings~\cite{resnet}. Our implementation of \sys\ in the image domain translates a human's simulated advice to the classifier in response to a LIME explanation expressed using superpixel features. Using this simulated advice, we demonstrate that this implementation significantly improves system accuracy, compared to a strong baseline, in a few shot setting. To accelerate future research, we are releasing our \sys\ image domain code at: \url{https://github.com/uw-hai/LIMEADE}.

To establish the generality of our approach, we perform a second case study in a very different domain with a different task. In this second case study, we incorporated \sys\ within \sanity, a publicly-deployed research paper recommender system with hundreds of users. While recommendations are made using an opaque neural model built on top of precomputed paper embeddings~\cite{specter2020cohan}, \sys\ allows humans to provide advice in terms of  unigrams and bigrams (\eg, marking them as of interest or not) that are suggested by an approximate, linear explanatory model.  In a simulation study based on organic user feeds in the log data, we show that explanation-based advice taking improves recommender quality, but we fail to find a significant improvement compared to adding a comparable number of labeled instances. We also perform an in-person user study showing that users feel that the ability to provide high-level feedback significantly improves their sense of trust, control and system transparency. 

Moreover, our work reveals that some ways of soliciting user advice may cause tension between explanation quality and advice diversity, potentially limiting the user's ability to adjust the ML model. We observed this {\em explanation-action tradeoff} in our second case study, where constraints on the user interface allowed us to accept advice on just a small number of the potential explanation terms. Such advice created a feedback loop, powered by iterative applications of advice, that reduced explanation diversity and hence limited users' future opportunities to further improve the classifier.

Significantly, our paper leaves a number of questions surrounding the advice-taking problem unanswered. For example, we do not conclusively answer the question of whether advice-taking methods allow end-users to improve the accuracy of real-world, opaque models more easily than by  annotating more instances. Moreover, in our image domain experiment, we uncover that the effectiveness of advice-taking methods may decrease with more supervision. Lastly, it is important to further study how advice-taking fits into broader frameworks within human-in-the-loop machine learning that incorporate human interventions with design parameters, model and algorithm choice, error tolerance, and beyond \cite{amershi_power_2014}. In many ways, we view this paper as a ``Call to Action'' to galvanize more researchers to study the advice-taking problem for opaque machine learners, as it is a rich area of study within human-AI interaction with many questions still to answer.

\section{\sys: Advice Taking for Opaque Models}
\label{sec:framework}

In this section, we provide a formal overview of the \sys\ framework and detail how it can be applied to opaque machine learning models to enable advice taking. With \sys, we assume that the human would like to give advice to an opaque machine learning model. By {\em opaque}, we mean that the model architecture may be completely unknown, or (if known), it may have too many parameters and nonlinearities for a human to understand. However, we assume that the model's inputs and outputs are available and that the model can be retrained on new instances. We work in a semi-supervised learning setting, in which the goal is to learn a hypothesis that maps an $s$-dimensional real-valued input vector to a label (for classification) or a real-valued output score in $[-1,1]$ (\eg, for recommendation). We are given a set $\mathcal{X}_L$ of labeled training instances $(x, y, w)$, where $x \in \mathbb{R}^s$, $y$ is the value to be learned, and $w$ is the weight assigned to the instance when training. Additionally, we optionally have a large, dense pool $\mathcal{X}_U$ of unlabeled instances $(x)$. Our explainable machine learning problem setting closely follows that of previous work in explainable ML \cite{Ribeiro_LIME_2016,lundberg_2017}. We assume that each instance $x$ can be represented as a binary-valued vector $x'$ that lies in an {\em interpretable} space. For example, in the text domain, the dimensions of $x$ might contain embeddings produced by a transformer, whereas the dimensions of $x'$ would correspond to interpretable features such as term frequency-inverse document frequency (TF-IDF) values for n-grams.\footnote{Term frequency-inverse document frequency, or TF-IDF, is a method of text featurization. Each document is featurized according to a fixed set of $n$-grams, where the feature value for the $n$-gram is given by the $n$-gram's frequency in the specific document in question, times a weight that puts more emphasis on terms that are less common across the corpus \cite{jones}.} In the image domain, the dimensions of $x$ would be pixels, while the dimensions of $x'$ might be superpixels~\cite{Ribeiro_LIME_2016} or fine-grained features~\cite{Akata2015EvaluationOO,Koh2020ConceptBM}.

Given an instance $x$ to explain, our approach uses an {\em explanatory model} $g$ in the interpretable space that locally approximates the opaque classifier $f$, i.e., $g(h'(z)) \approx f(z)$ for $z'$ nearby $x'$.  The model $g$ can be any interpretable model, such as a decision tree or linear model, produced using LIME or a comparable method.  We refer to the method that produces $g$ as {\sc Explain}$(f, x, h')$. 

\begin{figure*}
    \centering
     \includegraphics[width=.7\textwidth]{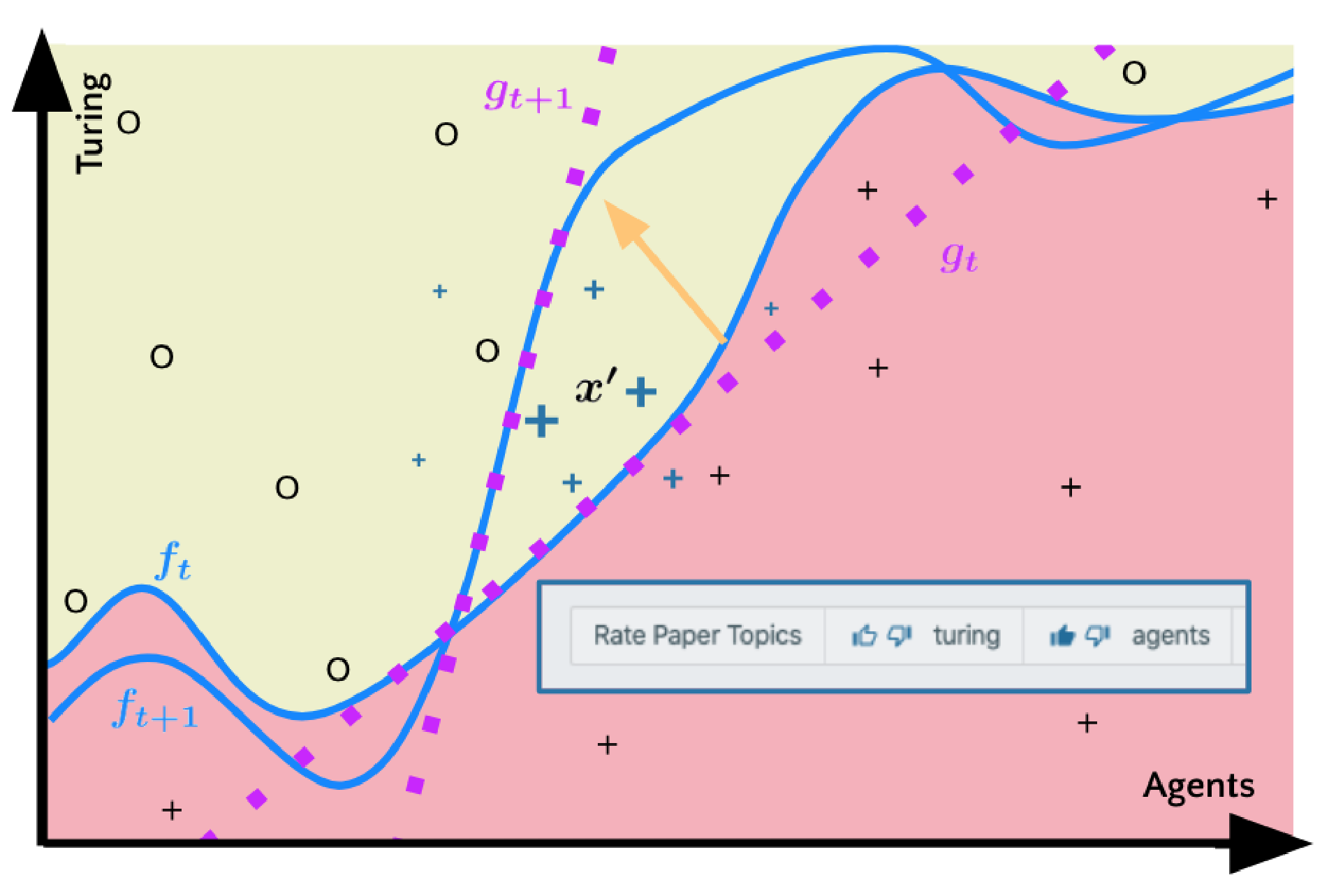}
  \caption{\sys\ updates an arbitrary opaque ML model by creating pseudo-instances. Here, we consider a recommender system for papers. Small black o's and +'s show the original training set (here, a user's ratings of papers), and shaded regions denote the complex boundary of the opaque classifier $f_t$. In order to explain a prediction, $h(x')$, the system generates a locally faithful explanatory model using {\sc LIME} or an alternative method.  This is $g_t$, shown as a purple dotted line. In practice, the explanatory model likely has many more than the two dimensions shown above, but suppose `Turing' and `agents' are highly weighted terms, hence used in the explanation. When the human specifies feature-level advice, e.g., `I want more papers about ``agents''', it could be used to directly alter a linear explanatory model (creating the new purple dotted line $g_{t+1}$); however, no simple update exists for an arbitrary, opaque classifier, which may be nonlinear and use completely different features, such as word embeddings. Instead, \sys\ generates positive pseudo-instances (shown as blue +'s) that have the acted-upon feature and are similar to the predicted instance. The pseudo-instances are weighted (shown by relative size) by their distance to the predicted instance $x'$ that was used to elicit feedback. By retraining on this augmented dataset, \sys\ produces an opaque classifier that has taken the advice, shown as a changed nonlinear decision boundary $f_{t+1}$.}
  \label{fig:pseudo}
\end{figure*}

Algorithm \ref{alg:framework} details \sys's approach to enabling a model to take advice, and Figure \ref{fig:pseudo} illustrates a concrete example of applying \sys\ on the paper recommendation domain. Given an instance of interest, $x$, we obtain an explanation $g(x')$ of the model's output $f(x)$ using {\sc Explain}$(f, x, h')$. The human can then provide a label on a feature of $x'$.  Informally, a positive label on feature $j$ of $x'$ represents the human's assessment that instances $z'$ near $x'$ should tend to be positive when $z'[j]=1$.  For example, a user of our paper recommendation system might give a positive label to the term ``BERT'' in a natural language processing paper to indicate interest in papers about the technique. Notably, a user's feedback is provided in terms of the high-level vocabulary of the explanatory model, not by eliciting new features to be added.

\newcommand{\R}{\mathbb{R}}

\renewcommand{\algorithmicrequire}{\textbf{Inputs:}}
\renewcommand{\algorithmicensure}{\hphantom}

\setlength{\textfloatsep}{0.1cm}
\setlength{\floatsep}{0.1cm}

    \begin{algorithm}[H]
    \begin{algorithmic}[1]
    \REQUIRE \hphantom{hi} \leavevmode \\
    $\mathcal{X}_L, \mathcal{X}_U$ \hphantom{aaaaaaaaaaaaaaaaaa}  // sets of labeled and unlabeled instances \leavevmode \\
    $f_t: \mathbb{R}^s \to [-1, 1]$  \hphantom{aaaaaaaaa} // opaque classifier, version at time $t$ \leavevmode \\
    $x \in \mathbb{R}^s$, $x' \in \{0, 1\}^{s'}$ \hphantom{aaaaaaa} 
    // instance \&\ instance in interpret. rep. \leavevmode \\
    
    $h':\mathbb{R}^s \to \{0, 1\}^{s'}$ \hphantom{aaaaaaaaa} // mapping s.t. $x' = h'(x)$ \leavevmode \\
    
   $\pi_{x'}: \{0, 1\}^{s} \to \mathbb{R}_+$ \hphantom{aaaaaaaaa} // weighting based on distance \leavevmode \\    
    $k \in \mathbb{N}$ \hphantom{aaaaaaaaaaaaaaaaaaa} // number of pseudo-examples
    \hphantom{a} \leavevmode \\
    \STATE $g_t =$ \textsc{Explain}($f_t$, $x$, $h'$) \hphantom{aaaaa} // obtain explanatory model \leavevmode \\ %
    \STATE 
    \textsc{Display}($g_t$, $x'$)  \hphantom{aaaaaaaaaaa} // display key features of $g_t(x')$ to end-user, \leavevmode \\
    \STATE  //  who then selects one feature (indexed $j$) as $+$ or $-$ indicator of instance \leavevmode \\
    \textbf{receive} $a \in \{-1, 1\}$ \AND $j \in \{1, \dots, s'\}$ 
    
    \STATE // select $k$ instances, label them using action $a$, and weight according to distance from $x'$
    \\ $\mathcal{N}_x \gets \{\}$
    \FOR{$1, \dots, k$}
        \STATE $\tilde{x} = $ \textsc{GetInstance}$(x, x', \mathcal{X}_U)$
        \IF{$h'(\tilde{x})[j] = 1$}
            \STATE $\mathcal{N}_x \gets \mathcal{N}_x \cup \{(\tilde{x}, a, \pi_{x'}(h'(\tilde{x})))\}$
        \ENDIF
    \ENDFOR    
    \STATE $\mathcal{X}_L \gets \mathcal{X}_L \cup$ $\mathcal{N}_x$

    \STATE $f_{t+1} \gets$ \textsc{Retrain}($\mathcal{X}_L$, $f_t$)

    \RETURN $f_{t+1}$
    
    \end{algorithmic}
     \caption{\textbf{Enabling an opaque model to take advice using \sys.} Given a set of required inputs, \sys\ solicits human advice in response to an explanation of a classified instance and retrains the opaque model accordingly. \textsc{Explain} is a function that generates an explanation for a given model and instance. }\label{alg:framework}
    \end{algorithm}

\sys\ uses the human's action to improve the opaque model $f$ by creating a set of $k$ training pseudo-instances with repeated calls to  {\sc GetInstance}$(x, x', \mathcal{X}_U)$. We experiment with two implementations of {\sc GetInstance}: sampling and generative.  Sampling  from the unlabeled pool is effective when the unlabeled pool is relatively dense, meaning one can acquire many instances with interpretable features similar to those of $x'$.  Generative approaches can be helpful when data are less dense.  For example, with images, \sys\ can create synthetic pseudo-instances by greying out random subsets of the superpixels in the input image, essentially reversing LIME's process for generating the explanatory model, $g$. The generative approach also works in the textual domain, \eg, by creating a synthetic document with nothing but the tokens selected by the user.

\sys\ only retains the pseudo-instances that contain the acted-upon feature $j$, i.e. those $\tilde{x}$ for which $h'(\tilde{x})[j]=1$.  \sys\ then assigns a value to each pseudo-instance according to the user action: $+1$ if the user assigned a positive feature label, and $-1$ otherwise. 

\sys\ assigns each pseudo-instance a weight based on its proximity to $x'$, with instances more similar to $x'$ given higher weight.\footnote{We measure proximity in the interpretable space, but it is equally possible to measure in the original space instead.} The reasons to weight local instances more highly are twofold: the explanatory method may only be locally correct \cite{Ribeiro_LIME_2016}, and the human actions may only be locally applicable.  For example, the positive label on ``BERT'' discussed earlier is helpful within the local scope of natural language processing papers, but could become misleading if applied globally---in biology papers for example, the term ``BERT'' often refers to a different meaning (the ``BERT gene'').  After selecting and weighting the pseudo-instances, \sys\ can optionally condense the selections (e.g., collapsing the instances into a single centroid).  Finally, \sys\ adds the resulting pseudo-instances to the labeled training set $\mathcal{X}_L$ and calls {\sc Retrain} to train the classifier $f$ on the new data set.  

While Algorithm \ref{alg:framework} is written in terms of binary classification, our approach generalizes naturally to the multiclass setting. This would entail that step 3 in Algorithm \ref{alg:framework} solicit not only which feature was a positive or negative indicator, but also for which class---pseudo-instances would then be labeled in step 8 with respect to the chosen class.  In the case of negative indicators in the multiclass setting, the pseudo-instances could be assigned random classes other than the chosen class.

We reiterate that \sys\ is general in many senses: our framework is model-agnostic, applies to a diverse range of advice-taking applications, and enables different forms of advice taking. In the next sections, we present two case studies that highlight the general applicability of \sys.

 \begin{figure*}
  \centering
\includegraphics[width=0.85\linewidth]{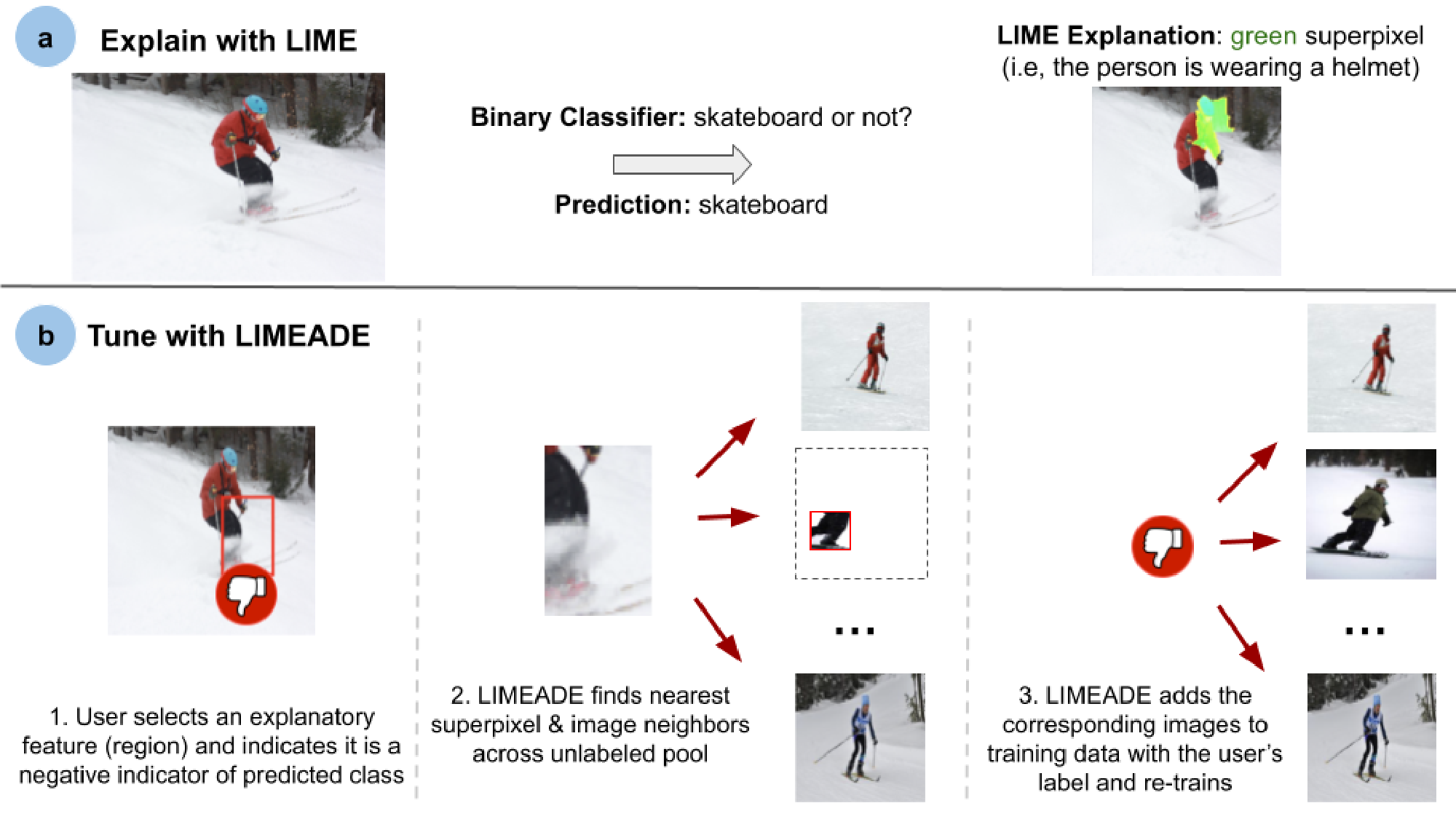}
\vspace*{-.2in}
  \caption{
 \textbf{a)} Suppose an opaque classifier incorrectly classifies an image of a skier as a positive instance of skateboarding. Suppose further that LIME returns an explanation showing a superpixel containing the skier's helmet as a positive indicator of the skateboarding class. Having seen this explanation, the user realizes that the classifier is predicting ``skateboard'' based on a spurious confound and should be looking elsewhere (we note that the end-user, such as a crowd worker, must understand the classification task but needs neither domain-specific knowledge nor an understanding of machine learning). \textbf{b)} While a helmet is an appropriate positive indicator for the skateboarding class, the user gives the advice that another superpixel, containing skis and ski poles, is a negative indicator. \sys\ translates this advice by updating the opaque model and retrieving unlabeled images and superpixels most similar to this ski superpixel (in our experiment, we retrieve the 50 most similar). The corresponding full images are then added to the training data --- with negative labels --- and the model is retrained, completing the \sys\ update. In general, a false positive classification will lead to negative feedback, and a false negative classification will lead to positive feedback (as illustrated in Figure \ref{fig:pseudo}).
 }
  \label{fig:imagedomain}
\end{figure*}

\section{Case Study 1: \sys\ for Image Classification}\label{sec:image}
We now present our evaluation of \sys\ in the image domain in order to study whether \sys\ allows humans to update real-world models more effectively than simply labeling more instances. In particular, we use \sys\ to enable updates based on simulated end-user advice for twenty deep neural image classifiers, \eg, a skateboard detector or fire hydrant detector. In Figure \ref{fig:imagedomain}, we illustrate an example of how \sys\ is used to process high-level advice in this context. With this simulated image domain experiment, we wanted to study the following research questions:
\begin{enumerate}
    \item Does advice taking with \sys\ further improve classifier performance as compared to a rigorous baseline of adding more labeled instances?
    \item How do \sys-powered improvements change as a function of supervision?
\end{enumerate}

\begin{table*}
\centering
\begin{tabular}{cccccl}
    \toprule
     \textbf{Class} & \textbf{2-Shot Accuracy} & \textbf{$\Delta$ Baseline} &  \textbf{$\Delta$ \sys} & \textbf{p-value} & \textbf{Winner} \\
    \midrule
    Baseball Glove & 76.73\% & 10.64\% $\pm$ 1.25\% & 10.75\% $\pm$ 1.52\% & 0.91 & \sys \\
    Snowboard & 75.97\% & 10.67\% $\pm$ 1.08\% & 10.74\% $\pm$ 1.42\% & 0.93 & \sys \\
    Giraffe & 74.54\% & 12.30\% $\pm$ 1.28\% & 16.42\% $\pm$ 1.51\% & $9.3 \times 10^{-8}$ & \textbf{\sys *}\\
    Carrot & 72.82\% & 7.37\% $\pm$ 1.15\% & 9.76\% $\pm$ 1.39\% & $4.9 \times 10^{-4}$ & \textbf{\sys *}\\
    Surfboard & 72.59\% & 8.23\% $\pm$ 1.19\% & 8.96\% $\pm$ 1.36\% & 0.34 & \sys \\
    Fork & 72.14\% & 7.70\% $\pm$ 1.24\% & 6.96\% $\pm$ 1.64\% & 0.53 & Baseline \\
    Sink & 71.55\% & 10.38\% $\pm$ 1.12\% & 10.44\% $\pm$ 1.49\% & 0.95 & \sys \\
    Cow & 69.90\% & 8.66\% $\pm$ 1.00\% & 11.53\% $\pm$ 1.07\% & $5.5 \times 10^{-6}$ & \textbf{\sys *}\\
    Donut & 67.51\% & 8.23\% $\pm$ 0.96\% & 9.65\% $\pm$ 1.11\% & 0.093 & \sys  \\
    Toothbrush & 65.85\% & 5.26\% $\pm$ 0.82\% & 4.86\% $\pm$ 1.02\% & 0.63 & Baseline \\
    Knife & 65.47\% & 7.31\%$\pm$ 1.10\% & 7.43\% $\pm$ 1.43\% & 0.86 & \sys \\
    Bed & 65.16\% & 9.50\% $\pm$ 0.93\% & 11.73\% $\pm$ 1.16\% & $5.7 \times 10^{-3}$ & \textbf{\sys *} \\
    Horse & 63.66\% & 8.49\% $\pm$ 0.90\% & 10.20\% $\pm$ 1.31\% & 0.050 & \textbf{\sys *} \\
    Cake & 63.48\% & 8.53\% $\pm$ 1.02\% & 9.07\% $\pm$ 1.33\% & 0.49 & \sys  \\
    Motorcycle & 63.16\% & 9.37\% $\pm$ 0.98\% & 15.97\% $\pm$ 1.08\% & $3.7 \times 10^{-11}$ & \textbf{\sys *} \\
    Frisbee & 62.67\% & 7.80\% $\pm$ 0.85\% & 7.03\% $\pm$ 1.18\% & 0.32 & Baseline \\
    Skateboard & 61.09\% & 6.93\% $\pm$ 0.82\% & 7.52\% $\pm$ 1.03\% & 0.48 & \sys \\
    Fire Hydrant & 59.21\% & 6.31\% $\pm$ 0.75\% & 8.38\% $\pm$ 0.92\% & $8.7 \times 10^{-3}$ & \textbf{\sys *} \\
    Scissors & 57.38\% & 6.51\% $\pm$ 0.79\% & 4.92\%  $\pm$ 1.05\% & 0.037 & \textbf{Baseline*}\\
    Suitcase & 55.65\% & 4.04\% $\pm$ 0.63\% & 4.23\% $\pm$ 0.67\% & 0.71 & \sys \\ \hline
    Total & 66.83\% & 8.21\% & 9.33\% & $2.3 \times 10^{-9}$ & \textbf{\sys *} \\
  \bottomrule
\end{tabular}
\caption{Updates using \sys\ boost the accuracy of an opaque image classifier more than the baseline. Results are shown for 20 classes averaged over 100 randomly-initialized runs each, and the accuracy boosts are reported relative to an average initial, 2-shot accuracy on a test set. For the updates, standard errors dare reported, and a $*$ indicates $p$-value $\le 0.05$. \sys\ outperforms the baseline on 16 of 20 classes and provides an overall boost of 9.33\%, as opposed to the baseline's overall boost of 8.21\%.}
\label{tab:limeadeimage}
\end{table*}
 
\subsection{Experimental Setup}
In order to determine whether \sys\ can support advice taking in the image domain, we evaluated on binary image classifiers, each comprising a logistic regression model trained on pre-computed image embeddings. As a base image dataset, we utilized $20,000$ images from the COCO dataset \cite{coco}. In order to create superpixel features for \sys\ feedback, we leveraged the same segmentation algorithm \cite{scikit-learn} used by LIME to compute superpixels for all 20,000 images. To generate embeddings for all full images and corresponding individual superpixels, we retrieved their representations from the penultimate layer of a ResNet-50 backbone pre-trained on ImageNet \cite{resnet, imagenet}. For a given superpixel, we computed the corresponding embedding by feeding the mimimum bounding box containing the superpixel to the embedding model. Pre-computing these embeddings resulted in a bank of embeddings for 20,000 images along with embeddings for all corresponding individual superpixels.

In order to ensure that our embeddings had not already been trained on the target classes in our experiment, we tested binary classifiers only on all 20 classes that are in COCO but not in ImageNet-1000.\footnote{The 20 classes are: baseball glove, snowboard, giraffe, carrot, surfboard, fork, sink, cow, donut, toothbrush, knife, bed, horse, cake, motorcycle, frisbee, skateboard, fire hydrant, scissors, and suitcase.} We wanted to measure the performance of a \sys\ update relative to a baseline update, so we completed 100 randomized initial configurations for each class. Moreover, for each configuration, we randomly constructed an initial training set of one positive and one negative instance (experiments in the 10-shot setting were less-effective, as described in Section \ref{sec:improvingquality}). We evaluated the two-shot accuracy of a logistic regression model on a held-out validation set and then performed one of the following two updates with both a randomly-drawn positive instance and a randomly-drawn negative instance simultaneously to preserve class balance:

\begin{enumerate}
    \item \underline{Baseline}: We update the model by adding the positive and negative instances to the training data and retraining.
    \item \underline{\sys}: First, we generate LIME explanations of the opaque classifier for both the positive and negative instance. In the positive case, we simulate a human's advice in response to the explanation by utilizing the COCO segmentation masks to automatedly give the superpixel(s) indicative of the class a positive label (i.e., in the case of ``giraffe,'' we select all superpixels containing giraffes using the COCO segmentation masks in the image labeled as ``giraffe''). In the negative case, we give the superpixel most influencing the LIME explanation a negative label. We then generate embeddings of these labeled regions and use the embeddings to retrieve the nearest superpixels and full images across the unused pool (consisting of $19,996$ images, along with their individual superpixels). We append the embeddings of the nearest neighbors' corresponding full images to the training data along with $+$ and $-$ labels, respectively, and retrain. This simulated approach to human advice enabled us to study the effectiveness of \sys\ updates by testing many initial configurations across a range of image classes. 
\end{enumerate}

We wanted to evaluate LIMEADE across different hyperparameter settings, so we varied the number of nearest neighbors included in the update ($n_{neighbors} = \{1, 5, 10, 25, 50, 100\}$), as well as the relative sample weight of the update ($w = \{0.25, 0.5, 1, 2, 4\}$), and performed a grid search. We evaluated performance on a balanced, held-out validation set of 400 positive instances and 400 negative instances for each class and selected the hyperparameters with the highest validation accuracy. This process yielded a relative sample weight of $0.25$, as well as $50$ nearest neighbors included in the update. With these hyperparameters selected, we then evaluated final performance on a separate, held-out test set of 400 positive instances and 400 negative instances for each class.

\subsection{\sys\ Feedback is More Effective than the Baseline}

In Table \ref{tab:limeadeimage}, we report the net changes in classifier accuracy when making updates with \sys\ and the baseline across all 20 classes and 100 runs per class, as evaluated on the test set. We find that \sys\ updates with simulated advice outperform the baseline for 16 of 20 classes, giving an average boost of $9.33\%$ compared to the baseline's average boost of $8.21\%$. These results are statistically significant: a paired t-test of \sys\ against the baseline yields a $p$-value of $2.3\times 10^{-9}$ across all 2,000 runs.

\subsection{Diminishing Returns as Supervision Increases}\label{sec:entanglement}
While conducting our case study with the image domain, our empirical results indicated that the \sys-powered improvements decrease as a function of more supervision. For example, repeating the experiments in the 10-shot setting, we find that \sys\ gives an overall boost of 0.63\%, whereas the baseline gives an overall boost of 0.88\%. It is important to note, however, that there is a fundamental entanglement between training data and supervision with respect to \sys. \sys\ is most valuable when the original supervisory data is subject to spurious correlations (i.e., when teaching ``cat,'' if all cats seen in the training data happen to be black, a \sys\ update communicating that color does not matter has high utility). If the training data is representative (because the training data contains more instances or because the instances themselves are better-selected), we expect a \sys\ update to provide less utility, as there are fewer potential spurious correlations for a human to correct via a \sys\ update. Indeed, as the quality of the originally-trained classifier approaches perfection, the value of \sys\ goes to 0, much as the  value of more training data also decreases. Our empirical evidence thus agrees with our intuition that a \sys\ update is most valuable in the low supervision setting.

\section{Case Study 2: \sys\ for Paper  Recommendation}
\label{sec:system}

For our second domain, we selected text ranking both for variety and importance. The overwhelming influx of new scientific publications poses a daily challenge for researchers \cite{bhagavatula-etal-2018-content,ekstrand_recsys,he_2010,Kanakia_MAG_2019,Sinha_MAG_2015}. However, based on Beel~\textit{et~al.}~\shortcite{beel_research-paper_2016}'s survey of 185 publications on academic paper recommendation, only a few systems explain why papers have been recommended or respond to user feedback other than liking/disliking specific papers, and all such systems rely on interpretable recommenders \cite{bakalov_2013,kangasraasio_2015,bruns_what_2015,verbert_2013,parra_user-controllable_2015}. The ability to explain and take advice for higher-performance paper recommenders, therefore, fills an important void.

Furthermore, a complete evaluation of a human-AI interaction approach requires testing it with real users in the loop \cite{Amershi2019GuidelinesFH}. For \sys, we  wanted  human users who were authentically motivated to understand and improve an ML classifier. In this regard, we built \sanity, a computer science (CS) research-paper recommender system based on Andrej Karpathy's arXiv Sanity Preserver \shortcite{karpathy_2015}. Deployed as a publicly-available platform, \sanity\ enables users to curate feeds from over 150,000 CS papers recently published on arXiv.org. With this testbed, users are implicitly incentivized to understand and improve the recommender system powering their feed in order to receive more interesting papers. Note further that each user is a task expert, since the users determine their own preferences.

Lastly, this study complements the first case study presented in the previous section. We intentionally selected two case studies with very different settings: while our first case study considered image classification, this study surrounding text ranking considers a different domain and task. Thus, studying and evaluating our implementation of \sys\ with \sanity\ provides evidence for the generality of our framework.

\subsection{Neural Recommender}\label{sec:neural}
To generate individual recommendations, we utilize a neural model consisting of a linear SVM on top of neural paper embeddings pre-trained on a similar papers task \cite{cohan2020}. Each paper is represented by the first vector (i.e., the \textsc{[CLS]} token typically chosen for text classification) after encoding the paper title and abstract using SciBERT \cite{beltagy-etal-2019-scibert}.  The neural embedding model is finetuned on a triplet loss $\mathcal{L} = max(0, v_i^T v_+ - v_i^T v_- + m)$ where $m$ is a margin hyperparameter and $v_i$, $v_+$ and $v_-$ are the vectors representing a query paper, a similar paper to the query paper, and a dissimilar paper to the query paper, respectively.  The similar paper triples are heuristically defined using citations from the \textsc{Semantic Scholar} corpus \cite{Ammar2018ConstructionOT}, treating cited papers as more similar than un-cited papers.  Recommendations are generated by training the model on a user's annotation history, with additional negative instances randomly drawn from the full corpus of unannotated papers.

\begin{figure*}
  \centering
\includegraphics[width=0.8\linewidth]{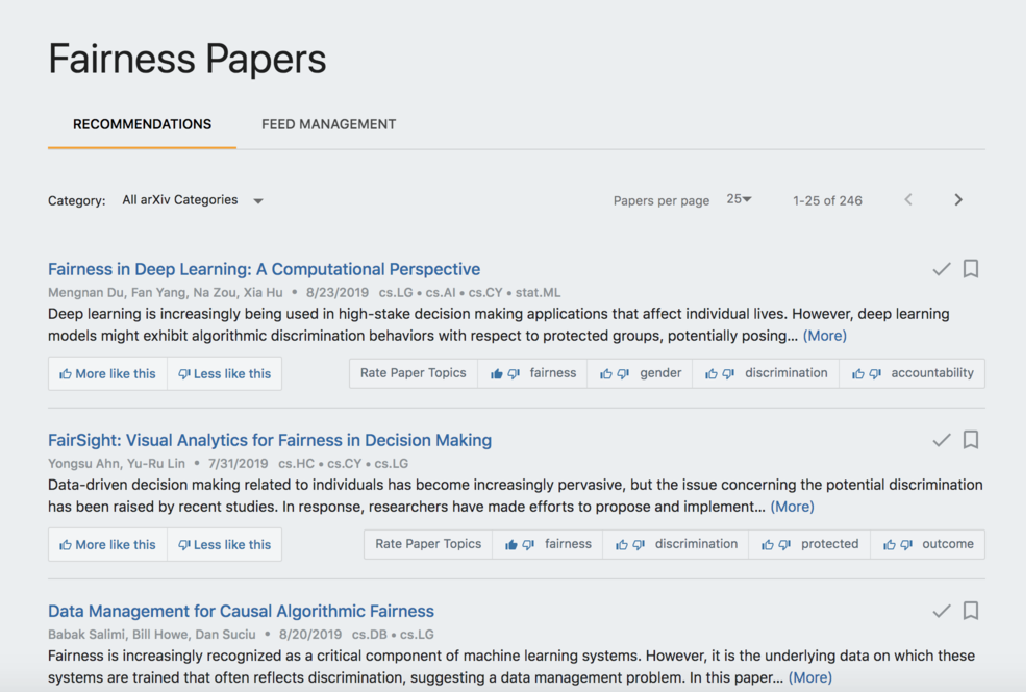}
  \caption{The UI for a feed in \sanity. Users can rate the papers themselves with the ``More like this'' and ``Less like this'' buttons, a standard feed affordance. Under each paper, the system also presents four terms to explain why it was recommended and solicits feedback with ``Rate Paper Topics'' --- by clicking thumbs up or down, the user can give advice by requesting that the feed include more or less of the specified topic.
  }
  \label{fig:UI}
\end{figure*}

A user begins the process of curating their feed by either selecting a specific arXiv CS category or issuing a keyword search and then rating a handful of the resulting papers.  A feed consists of a list of recommended papers sorted by predicted recommendation score (see Figure \ref{fig:UI}).  Each paper can be rated using traditional ``More like this'' or ``Less like this'' buttons underneath each paper description. 

\subsection{Implementation of Explanations and Feedback}\label{sec:implementation}

The UI for \sanity\ (Figure~\ref{fig:UI}) displays a list of recommended papers and adorns each with an explanation comprising four terms; to the left of each term are thumbs-up and thumbs-down buttons, enabling the user to not only rate the papers themselves but also {\em give advice} in response to the explanation and indicate if they would like to see more or fewer papers related to that term. We refined our user interface design through iterative informal user testing. The explanatory terms are generated using a simple, explanatory model (\sys's {\sc Explain} function), which we implement as a linear model over uni- and bigram features. In particular, our linear model is defined as $g(x')=w_0 + \sum_i w_i x'_i$,  and the explanation surfaced for $g(x')$ consists of high-impact terms in the model, i.e., those with high values for the product $w_i x'_i$. Specifically, we select the 20,000 features with the highest term frequency across our corpus. Our approach of using a post-hoc explanatory model is similar to that used by LIME, except to enable real-time performance  our explanatory model is trained as a global, rather than local, approximation of the neural model \cite{Ribeiro_LIME_2016}. This global approximation was chosen because testing on an early prototype revealed that generating explanations for a feed using LIME was too computationally expensive, since LIME requires sampling nearby instances and training a model for \textit{each} recommendation on the page; this latency negatively impacted the recommendation experience.\footnote{Note that in a LIME-style approach to instance-level explanation, one needs to either sample neighbors around the instance (which may be quite distant) or generate synthetic instances, which in our case requires generating paper embeddings on the fly. Both approaches introduced unacceptable latency.}

Given the explanatory model, \sys's {\sc Display} function chooses explanations to display by computing each term's contribution to the output of the linear model for the given paper, which is equal to the product of the term's TF-IDF value for the paper with the term's feature weight in the linear model. We note here that even though the explanatory model is a global approximation, the explanations are local ones, as this product encodes instance-specific information on why the paper has been recommended.\footnote{In a later implementation of Semantic Sanity, the platform also included a global explanation surfaced at the top of each feed, which is the subject of another study \cite{marissa_22}.} Next to each explanatory term are thumbs up and down buttons (see Figure~\ref{fig:UI}). When the user provides feedback with these buttons, \sys\ generates pseudo-instances and retrains the neural recommender. We use a generative approach within {\sc GetInstance} that leverages the unlabeled pool of papers.  We select the top 100 papers from the full corpus with the highest TF-IDF value for the feedback term and generate a single synthetic pseudo-instance (i.e., we use $k=1$) equal to the centroid of these papers' embeddings with a weight of $1$. The instance is appended to the user's history and labeled with the user's annotation of the term (+/-). 

\subsection{Online Traffic}

In the next section, we describe a controlled user study comparing \sanity\ with and without \sys. However, since its public launch, \sanity\ has also attracted considerable organic traffic: users with accounts have constructed 2,478 feeds and have logged 21,713 paper annotations and 1,320 topic annotations (we note that annotating topics was only possible after the \sys-based implementation was introduced on November 11, 2019, five months after the initial launch of \sanity). The target user base was computer science researchers, and the platform was advertised through social media and email lists. In Sections \ref{sec:log-study} and \ref{sec:diversity}, we analyze a subset of the organic user logs as a complementary part of our evaluation.

\subsection{User Study: Experimental Setup}

In order to evaluate the effectiveness of our \sys-based system for recommending papers with real users, we performed an in-person user study. With this user study, we wanted to address the following research questions:
\begin{enumerate}
    \item Do participants prefer \sys\ over a baseline of just explanations according to self-reported ratings of trust, control, transparency, intuition, paper coverage, and the overall system? 
    \item Does \sys\ increase participants' feed quality, evaluated quantitatively with blind ratings of recommended papers?
    \item How do participants utilize the topic-rating affordances powered by \sys? \item What constructive feedback do participants have surrounding our particular instantiation of \sys\ with \sanity?
\end{enumerate}

We recruited 21 participants through a public university's computer science email lists. All participants were adults who reported experience with reading computer science research papers in our screening questionnaire. Each session lasted one hour, and each participant was compensated with a \$25 Amazon gift card. Our IRB application did not include a plan to collect and share participants’ demographic data, and therefore, we could not include it in the study results.

Participants were asked to curate feeds of computer science papers pertaining to a topic of their choice using two different recommendation user interfaces (UIs), one that used \sys\ to provide advice-taking explanations, and one that did not present explanations, instead only allowing users to rate the papers themselves (the baseline); other than this difference, the UIs were the same.  The participants were asked to choose a topic that they were interested in following over time as new papers are added to the arXiv, but not so general that it is already covered by an existing arXiv CS category (e.g., artificial intelligence). Once a topic was selected, each participant was asked to name the desired feed, which served as the goal for curation using both UIs.  

Each participant began curation by selecting exactly three seed papers that were then used to initialize the feeds in both UIs.  Both systems surfaced the same initial recommendations in response to the participant's three seed papers and thus had identical initial states. Each participant was then presented with one of the two UIs and given instructions on how to use it. 11 participants received the baseline system first, and 10 received the \sys\ system first. They were then presented with the second UI. For both UIs, the participants were told to use as many or as few annotations as desired until their feed was curated to their liking, or a maximum of 10 minutes was reached. We recorded the participants' annotations for both feeds. After using each UI, the participants were asked to complete a short survey. They were then asked to rate a blind list of combined recommendations from the two feeds that they had curated, according to whether they would like to see each paper in their desired feed.  These recommendations were generated on a held-out paper corpus, disjoint from the papers available within the feed UIs. 

Data were successfully collected for all 21 participants.  The participants' chosen topics varied greatly, including ``Spiking Neural Networks,'' ``Moderation of Online Communities,'' and ``Dialogue System Evaluation.'' 

\subsection{User Study: Quantitative Results}\label{sec:results}

\subsubsection{User Experience: Participants Prefer \sys}\label{sec:user_ratings}
In the surveys administered after using each UI, we asked each participant to provide overall ratings for each system and to state which system they preferred along dimensions such as trust and intuitiveness.  The results are summarized in Tables~\ref{tab:ratingsone} and \ref{tab:ratingstwo}.\footnote{For all statistical significance tests, we report adjusted $p$-values using the Holm-Bonferroni procedure for multiple comparisons \cite{holm79} in \textsc{R}'s \textsc{p.adjust}
library \cite{rlanguage}.} 

\begin{table}
\centering
      \begin{tabular}{cccl}
        \toprule
        Which system... & Baseline & \sys & $p$-value\\
        \midrule
        ...trust more?& 4 &\textbf{17*}&0.043\\
        ...more control?&0&\textbf{21*}&$\approx$0\\
        ...more transparent?&3&\textbf{18*}&0.012\\
        ...more intuitive?&\textbf{12}&9&0.664\\
        ...not missing relevant papers?&3&\textbf{18*}&0.012\\
      \bottomrule
    \end{tabular}
\caption{Among 21 participants, most prefer our system over the baseline when prompted with these questions. \textit{(*) indicates a statistically significant result under a two-sided binomial test against a null hypothesis of no preference between the systems}.}
\label{tab:ratingsone}
\end{table}

\begin{table}
\centering
\begin{tabular}{cccl}
    \toprule
     Likert scale rating & Baseline & \sys & $p$-value\\
    \midrule
    Overall system & 3.38 $\pm$ 0.59 & \textbf{3.85 $\pm$ 0.57}\textbf{*} & 0.043 \\
    Would use again? & 3.38 $\pm$ 1.16 & \textbf{3.90 $\pm$ 0.94} & 0.257 \\
  \bottomrule
\end{tabular}
\caption{Mean $\pm$ Standard Deviation of  21 participant ratings of each system. Ratings were on a scale from 1 (worst/no) to 5 (best/yes). \textit{(*) indicates a statistically significant result under a two-sided paired $t$-test against a null hypothesis of zero mean difference between the systems.}}
\label{tab:ratingstwo}
\end{table}

Overall, participants rated our approach significantly higher than the baseline.  They also rated it significantly higher on trust, control, and transparency, and on confidence that their recommendations were not missing relevant papers. Understandably, our \sys\ system appeared less intuitive to participants than the baseline due to the increased complexity of the UI, though this result was not statistically significant.  Finally, while not statistically significant due to small sample size, participants indicated more likelihood to use our system again over the baseline. In aggregate, these results indicate a higher-quality user experience with the \sys\ system than with the baseline system.

\subsubsection{Mixed Results for Feed Curation Time}\label{sec:curation_time}
In analyzing the times required by each participant to complete feed curation using the two systems, we observe that eight participants finished feed curation with the baseline system first, seven finished with the \sys\ system first, and the remaining six utilized all ten minutes for the curation of both systems.

\subsubsection{Most Participants used Both Paper and Topic Ratings}\label{sec:ratings}

\begin{figure*}[t!]
    \centering
    \begin{subfigure}[t]{0.5\textwidth}
        \centering
        \includegraphics[height=2.1in]{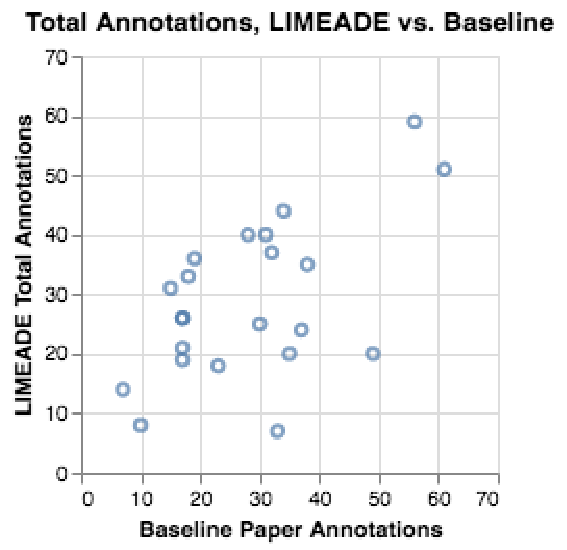}
    \end{subfigure}
        ~ 
    \begin{subfigure}[t]{0.5\textwidth}
        \centering
        \includegraphics[height=2.1in]{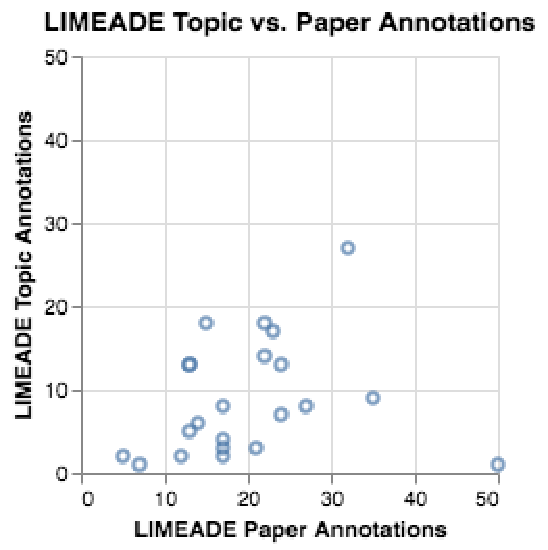}
    \end{subfigure}
    \caption{
    Scatter plots showing (left) the total number of annotations used to curate a feed with \sys\ (paper annotations + topic annotations) vs. the number of baseline paper annotations per user, and (right) the number of topic annotations vs. the number of paper annotations in the \sys\ system. Most participants used \sys-powered topic-level feedback as well as paper-level feedback.
    }~\label{fig:annotations}
\end{figure*}

To explore the breakdown of participants' rating habits with the baseline system and the \sys\ system, we present Figure \ref{fig:annotations}. In the left plot in Figure \ref{fig:annotations}, we observe that participants displayed a high degree of variance in the number of ratings applied during feed curation, ranging from 7 annotations to 61 annotations with the baseline system. Comparing the total number of annotations made using the system with \sys\ vs. the number of annotations made with the baseline, we find a best-fit slope of 0.913. This suggests that the participants made approximately the same number of annotations across both systems.

In the right plot, we observe that there is significant diversity in how participants applied topic annotations, ranging from 2 annotations to 27. However, most participants utilized a combination of paper and topic ratings, with more paper ratings than topic ratings on average. Interestingly, five out of the twenty-one participants provided more negative paper ratings than positive ones in the baseline; when presented with the \sys\ affordances, no participants provided more negative paper ratings than positive ones, but four participants applied more negative topic ratings than positive ones.

\subsubsection{Blind Ratings of Recommendations: No Significant Difference in Feed Quality}\label{sec:blindratings}

We also investigated whether the topic-level feedback provided by \sys\ measurably increased the quality of participants' feed. We showed participants the top 20 recommendations generated by both systems on the held-out corpus of papers and measured their ratings. Specifically, we computed the discounted cumulative gain (DCG)\footnote{We did not use NDCG because the participants liked different numbers of papers between the two systems.} and average precision (AP), common metrics for assessing recommendation feed quality. For DCG, we observe a mean difference of $0.259$ in favor of the baseline system recommendations; however, the corrected $p$-value for the two-sided, paired $t$-test for mean differences is $0.218$, indicating no significant difference in feed quality between the two systems under DCG. For AP, we observe a mean difference of $0.0412$ in favor of the baseline system, with a corrected $p$-value of $0.257$, also indicating lack of significant difference in quality under AP. Based on the constructive feedback that we received, we speculate that this result could be improved by making implementation-specific adjustments to \sanity.

\subsection{User Study: Qualitative Feedback}\label{sec:qualitative}

We analyze participants' text responses and provide a sample of quotes that complement the quantitative results. After using each system, participants were asked to provide free text responses to the question, ``Would you like to share anything else about using the system?'' At the end of the study, participants were also asked, ``Do you have any last thoughts that you would like to share regarding actionable explanations?'' Overall, participants found the advice-taking affordance granted by our system helpful: \textit{``The explanations here were especially useful in their capacity as decisions rather than just explanations. I would have found them really really annoying if they were presented only as an explanation of why you thought I would like a paper, rather than an attribute I could ask for more or less of.''} In particular, participants stressed the importance of the \sys\ affordance as a filtering mechanism: \textit{``The topics feature was excellent, because there are many papers which cover *some* topics I like but also some that I don't, and this let me pick that out.''}

The constructive feedback received in the qualitative responses illustrate a number of implementation-specific improvements that could be made to \sanity. The most common category of constructive feedback concerned the quality of terms in the explanations, mentioned by 10 participants. Though we utilized stemming to eliminate these redundancies in each paper explanation, we did not eliminate synonyms from the list of terms. For example, three of the ten participants specifically requested that abbreviations in explanations be removed or linked to full terms. These issues reflect the negative consequences of utilizing 20,000 TF-IDF terms for our explanatory model featurization. In addition, five of these users also stated that the terms were too general. We speculate that the term quality in the explanations negatively impacted the users' ability to give advice to the model via the LIMEADE affordance.

Similarly, three participants directly addressed what we term the explanation-action tradeoff in the next section, noting that the lack of diversity of terms in the explanations was limiting.  One participant commented: \textit{``After a few minutes, almost all the same terms that I had liked were coming up, so there were few new terms for me to thumbs up or down. I think if the system could focus on bringing up relevant papers that have a new term or two to which I can react, that would make the curation even better.''} This suggests tuning the system to favor more explanation diversity even more than we did in our initial implementation.

\begin{table*}[ht]
    \centering
      \begin{tabular}{cccl}
        \toprule
        \small{Number of labeled papers} & \small{Base Ranking Performance} & $\Delta$ \small{\sys}\ \\
        \midrule
        2 & 0.884 & \textbf{0.015} \\
        5 & 0.901 &\textbf{0.009} \\
        10 & 0.908 & \textbf{0.005} \\
      \bottomrule
    \end{tabular}
    \caption{Simulated evaluation of ranking performance  (NDCG) based on log data from actual usage in case study 2.  \sys\ improves performance over the baseline system, which does not use the annotated explanations.}
   \label{tab:simulation}
\end{table*}

Interestingly, two users believed that the set of topics surfaced was too restrictive, one thought that the terms were too diverse, and one thought the diversity was a good feature.  This provides some evidence that different users have different preferences for explanation diversity, suggesting that it should perhaps be tuned in a user-specific manner. Additionally, four participants commented on topic annotation strength, all of whom indicated that it was too potent, revealing the importance of empirically evaluating the optimal strength of an update. Based on this feedback, we reduced the annotation strength in our application following the evaluation.

\subsection{Feed Quality Revisited Using Log Data: Term Annotations in \sys\ can Improve Performance}
\label{sec:log-study}

We also investigated the effect of high-level advice on a different set of users --- those who used \sanity\ in the wild, rather than as part of laboratory a user study --- using the log data of the online deployment. Specifically, we compiled a data set of 1,636 rated papers across 30 feeds, where each feed had at least one annotated explanation (the average number of annotations for these feeds was 4.4 terms).  We evaluate two recommenders: a baseline ranker that uses only the rated papers, and a \sys\ ranker that uses both the rated papers and the annotated terms processed by \sys.  We evaluate at three different training sizes (2, 5, and 10 labeled papers), and to maximize the contrast between \sys\ and the baseline, we always provide \sys\ with {\em all} of the explanation annotations for the feed (4.4 terms per feed on average). Thus, this experiment measures whether \sys 's pseudo-instance approach can be effective given sufficiently informative term annotations, but is not an accurate simulation of the system in practice (in which term annotations would arise only from explanations on papers in the limited training set). For each feed and size we compute the average normalized discounted cumulative gain (NDCG) ranking performance for up to ten sampled training sets, testing on the remainder. The average of the NDCG statistics across feeds is our final evaluation measure.

Table \ref{tab:simulation} shows that \sys\ does improve performance over the baseline, but the benefits of the annotated explanations diminish as the number of initially rated papers increases.  The individual differences shown in the table are not statistically significant, but the aggregate performance over all three sizes shows \sys\ performing significantly better than the baseline (p-value 0.017, two-tailed paired t-test, after Holm-Bonferoni correction). \sys\ with 2 and 5 annotated papers performs comparably to the baseline with 5 and 10 annotated papers, respectively, meaning that \sys\ reduces the number of paper labels required to achieve a given level of performance by an amount roughly equal to its number of term annotations in this experiment.  The experiment is inconclusive regarding whether giving advice via term annotations would be {\em preferable} to obtaining a similar number of labeled instances in this domain. Experiments with more users and feeds are necessary to resolve these questions.

\section{Discussion}

\begin{table*}
\centering 
{\small
      \begin{tabular}{lll}
        \toprule
         & \multicolumn{1}{c}{\textbf{Case Study 1}} & \multicolumn{1}{c}{\textbf{Case Study 2}}\\
        \midrule
        \textbf{Domain} & Image & Text \\
        \midrule
        \textbf{Task} & Object detection & Paper recommendation\\
        \midrule
        \textbf{ML Setting} & Classification &Ranking\\
        \midrule
        \begin{tabular}{@{}l@{}}\textbf{Explanatory} \\ \textbf{Features}\end{tabular} & \begin{tabular}{@{}l@{}}Superpixels (${\approx} 10{-}25$),\\
        instance-specific
        \end{tabular}
        & \begin{tabular}{@{}l@{}}$n$-gram topics (3 shown of 20,000 possible), \\
        same vocabulary across instances \end{tabular}\\
        \midrule
        {\bf Feedback Mechanism}  &
        $\uparrow$ or $\downarrow$ vote any superpixel &
        \begin{tabular}{@{}l@{}l@{}} $\uparrow$ or $\downarrow$ vote any of 
        3 displayed n-grams \end{tabular} \\
           \midrule
        \begin{tabular}{@{}l@{}l@{}}\textbf{\sys\ {\sc GetInstance}} \\ \textbf{Method (for selecting } \\ \textbf{pseudo-instance)} \\
        \end{tabular} & \begin{tabular}{@{}l@{}l@{}} Retrieve nearest unlabeled  \\ images \& superpixels \end{tabular} & 
         \begin{tabular}{@{}l@{}l@{}} Generate a synthetic instance equal to the \\centroid of matching unlabeled instances \end{tabular} \\
        \midrule
        \textbf{Evaluation} & \begin{tabular}{@{}l@{}}Simulated study with clear \\ ground truth \end{tabular}  &  \begin{tabular}{@{}l@{}l@{}} In-person user study \& simulated study \\ using log data from online usage \end{tabular}  \\
        \midrule 
        \textbf{Results} & \begin{tabular}{@{}l@{}l@{}l@{}l@{}}Advice taking with LIMEADE \\ improves accuracy more quickly \\ than adding labeled instances, \\ though this effect  diminishes \\ with more initial training data. \end{tabular}  &  \begin{tabular}{@{}l@{}l@{}l@{}l@{}l@{}} Advice taking with LIMEADE leads \\ to increased control, trust, satisfaction, \\ \& system transparency. Inconclusive \\ w.r.t. improving accuracy more easily \\ than adding labeled instances. \end{tabular}  \\
      \bottomrule
    \end{tabular}
}
\caption{A comparison of our two case studies.}
\label{tab:casestudies}
\end{table*}

Evaluating on real-world domains with real human interactions is crucial in order to make progress in  human-centered AI broadly, and for advice taking in particular. This section considers broader questions of the connections between our two case studies, the effectiveness of human feedback, and the interactions between the fidelity of explanations and the affordances provided for action.

\subsection{Demonstrating the Generality of \sys}
\label{sec:improvingquality}

In order to demonstrate that \sys\ is a universal mechanism for applying high-level advice to an arbitrary ML model, we chose our case studies to span a diverse range of dimensions.
Table \ref{tab:casestudies} summarizes the differences, which include the source domain (image vs. text), type of model (classification vs. ranking), nature of the explanatory vocabulary, and method for generating pseudo-instances. There are many options for creating pseudo-instances, and future work will be necessary to uncover the best methods. For example, is it better to generate one instance (as we did in the text domain) or several (as we did in the image domain)? Is it better to label naturally occurring (unlabeled) instances as we did in the image domain, or to create a synthetic pseudo-instance as we did by computing the centroid of matching examples in the text domain?

Usage differed across the case studies as well. In the image domain, we evaluated the effect of a single piece of high-level advice on the accuracy of the classifier. In the text domain, however, users interacted repeatedly to improve the ranker by providing a sequence of high-level advice and labeled examples in the way that seemed most natural to them.

\subsection{When Does High-level Advice Improve Learner Accuracy?}
\label{sec:improvingquality}

When tested on numerous domains, we obtained positive to indeterminate results about the effectiveness of \sys\ processing high-level human advice. Does this reflect a weakness in the \sys\ approach or limitations of our LIME explanations? Or is it intrinsic --- perhaps human-interpretable vocabulary is simply too dissimilar to the features learned by modern neural methods for {\em any} human  advice to be useful. Maybe getting more data is the only or the most effective way for humans to help out?

One thing seems clear --- in order to answer this question, the research community must conduct more experiments on real world domains, rather than toy domains with artificial confounds, such as Decoy MNIST.

While they only simulate interactions, our image domain experiments (Section \ref{sec:image}) reflect actual human judgements about which regions contain the object in question. \sys-processed advice about which regions contained an object significantly improved classifier accuracy in the two shot case. However, when we conducted similar experiments after training the twenty classifiers with ten instances, we found no significant improvement. Perhaps this is because the model had already learned where the objects were located. More likely, it had found the context imparted from background information to be useful in the classification decision. It also may stem from the segmentation algorithm that induced the `advice vocabulary' or perhaps the \sys\ method weighted instances incorrectly.

While users clearly liked the ability to provide high-level advice and felt it increased their sense of trust and control, we found mixed results with respect to improving ranking accuracy as measured with DCG. Our controlled study over 21 users (Section \ref{sec:blindratings}) found no significant difference between feed accuracy incorporating \sys\ advice {\em vs.} feeds created with simple labeled instances. In contrast, we did find significant improvements stemming from \sys\ advice in our simulated log study on 30 different users (Section~\ref{sec:log-study}). The differences could stem from our \sys\ mechanism, the bi-gram vocabulary chosen as features in the explanatory model, the size of our study, or some other reason.

We strongly believe that much more research should be devoted to this important question. \sys\ is an important first step, but our paper should be considered a ``Call to Action'' for more investigation. To this end, we will release the code for \sys\ and our image experiments, including our modified COCO dataset with the precomputed superpixel vocabulary and corresponding embeddings.

\begin{figure*}
  \centering
  \includegraphics[width=0.65\linewidth]{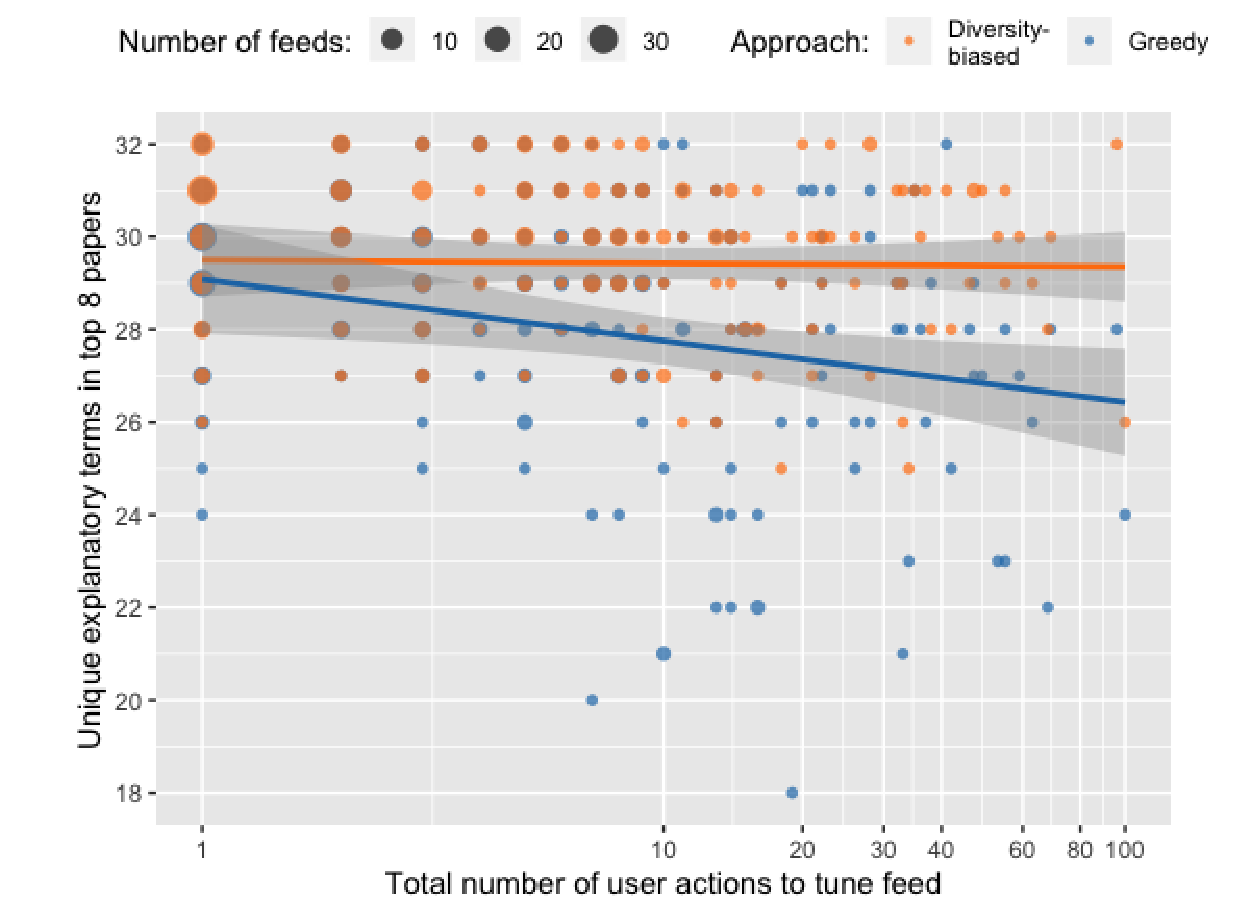}
  \caption{A scatter plot showing the number of unique explanation terms in the first page of the feed vs. the number of actions taken by the user in order to give advice to their their feed. Orange dots correspond to diversity-biased explanations currently used in the system. Blue dots correspond to greedy explanations, where the most important terms are surfaced without stochasticity. The size of each dot corresponds to the number of feeds in that bin. Note that greedy explanations (blue) display a stronger negative correlation between unique terms and term annotations than diversity-biased explanations (orange). Thus, the greedy approach limits opportunities for advice taking with topics as the feed curation process evolves, while the diversity-biased approach continues to facilitate advice taking with topics. 
  }\label{fig:tradeoff}
\end{figure*}

It is also important to contextualize our findings within prior work on advice taking. While studies such as \cite{kulesza_principles_2015} demonstrated a clear improvement in classifier accuracy in the setting of explanatory debugging, other studies have found the opposite. Of particular interest are the results from \cite{ahn_2007} and \cite{waern_user_2004}, which demonstrated that tunability for search and recommendation tasks can negatively impact feed quality when it takes the form of adding or removing terms from the featurization. Likewise, \cite{das_2013} shows that advice taking with interpretable models can lower accuracy. Lastly, \cite{sherry_tochi} is another datapoint that indicates that letting people into the interactive machine learning loop can be problematic. These concerns are especially problematic, given users' clear expectations that feedback will lead to ML improvement~\cite{SmithRenner2020NoEW}. For this reason, we reiterate that our paper is a ``Call to Action'' for more research surrounding high-level advice and learner accuracy.

\subsection{Exposing the Explanation-Action Tradeoff}\label{sec:diversity} 
\sanity\ chooses explanations to display by computing each term's contribution to the output of the linear model for the given paper, which is equal to the product of the term's TF-IDF value for the paper with the term's feature weight in the linear model.  The natural choice is to surface the terms with the highest-magnitude contributions in the linear model \cite{Ribeiro_LIME_2016}; we call this a \textit{greedy} approach. Users could then react to these presented terms, thumbing them up or down.

Comments from early users of our paper recommender indicated that there is a tradeoff between using the greedy explanation approach and the explanatory terms' uses as affordances for feedback, which we call the \textit{explanation-action tradeoff}. In particular, user action on an explanatory feature will lead the model to place increasing importance on it and correlated features.  With the greedy approach, these terms will begin to dominate both the model and the explanations, limiting the number of unique explanation terms and thus subsequent pportunities for feedback. For example, `thumbs-up'ing the term ``fairness'' causes papers about fairness to rise in the feed; under the greedy approach, these papers will contain the term ``fairness'' in their explanations, thereby crowding out new terms for the user to act on. 

Based on the feedback we received, our final implementation of {\sc Display} in \sanity\ uses a diversity-biased approach that samples explanatory features, in a way that prevents previously suggested terms from dominating subsequent explanations.\footnote{Specifically, {\sc Display} uses a parameter $\gamma$ and samples terms proportionally to the magnitude of term contribution, raised to the $\gamma$ power (higher values of $\gamma$ result in a more greedy approach; lower values increase diversity). We selected $\gamma = 4$ for our implementation. To further reduce term redundancy in each recommended paper's explanation, we used the Python \texttt{NLTK} PorterStemmer \cite{nltk} to deduplicate terms with the same stems (e.g., ``fair'' and ``fairness'') from each explanation.} However, as noted in Section \ref{sec:qualitative}, three participants commented in their qualitative feedback that they would have still liked even more diversity --- further evidence that properly considering and calibrating the explanation-action tradeoff is important for advice taking.

To illustrate the impact of the explanation-action tradeoff and the distinction between our diversity-biased approach and the canonical greedy method, we perform an analysis on the logs of 300 users' feeds from \sanity's online deployment. For each user, we compute \textit{(i)} the total number of actions the user has taken on displayed explanatory terms, and \textit{(ii)} the number of unique explanation terms among the latest top eight recommended papers under our diversity-biased \textsc{Display} implementation.  We then repeat \textit{(ii)} but with \textsc{Display} with $\gamma = \infty$ to simulate what explanatory terms the users would see today under a greedy approach.

\begin{figure*}[t!]
  \centering
  \includegraphics[width=0.75\linewidth]{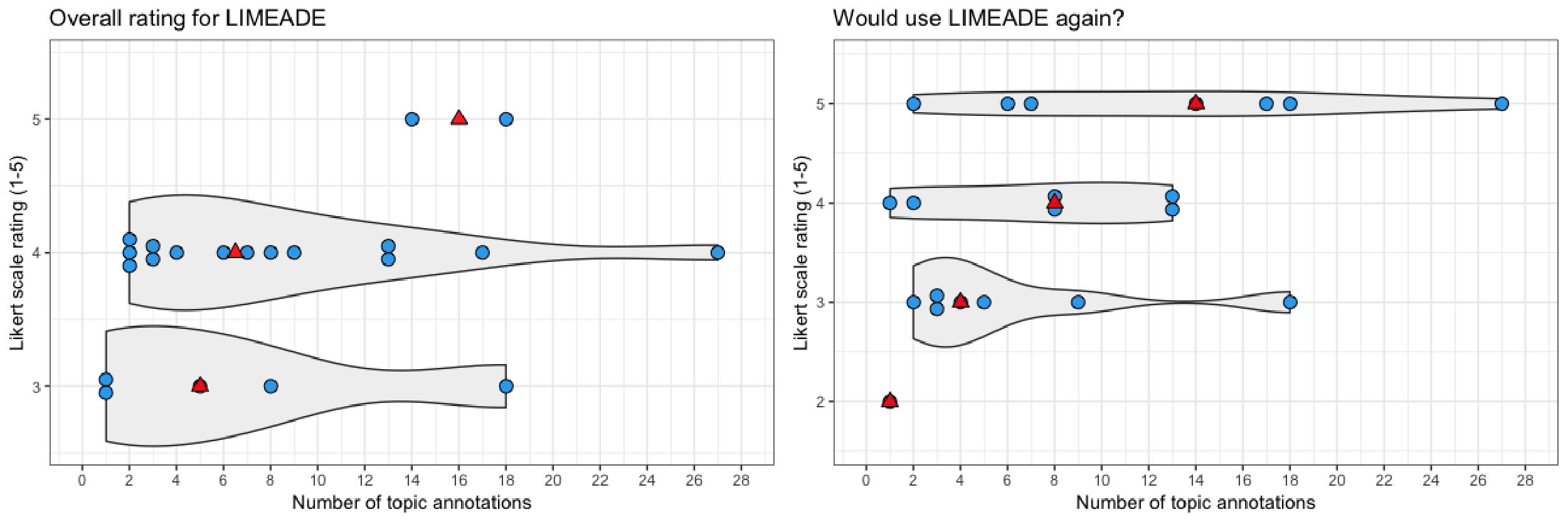}
  \caption{Plots showing participants' Likert scale evaluations of our overall \sys\ system (left) and the likelihood that they would use our system in the future (right) as functions of the number of topic annotations made when using our \sys\ system. The red triangles show the median number of annotations for each rating level. 
  }\label{fig:topic_ratings}
\end{figure*}

In accordance with the explanation-action tradeoff, we observe in Figure \ref{fig:tradeoff} that the number of unique explanation terms (i.e. advice-taking affordances) tends to be lower under a greedy approach. Furthermore, this effect grows stronger as users give advice to their feeds to be increasingly specific to a particular topic.\footnote{Figure~\ref{fig:tradeoff} likely understates the impact of the tradeoff, as users had been exposed to explanations under the diversity-biased approach prior to this analysis. Had they been exposed to explanations under the greedy approach for their entire sessions, we likely would observe an even stronger crowding-out effect.}
In contrast, the number of affordances remains relatively constant under our diversity-biased approach. Though some explanation terms with lower contribution weight are included within the explanatory model, our diversity-biased approach thus successfully mitigates the crowding effect observed with the greedy approach.

The explanation-action tradeoff is related to, but distinct from, the classical {\em explore-exploit tradeoff} faced by recommender systems and other machine learners \cite{sutton_reinforcement_2018}. The explore-exploit tradeoff entails deliberately passing up a known reward in the hopes of learning more about the reward structure in order to have better long-term gains. Thus, the explore-exploit tradeoff encourages taking a chance in executing an action in the hopes that it will provide a big reward, leading to frequent execution of the action in the future. The explanation-action tradeoff is similar to the explore-exploit tradeoff, in the sense that it entails deliberately declining to provide the most accurate explanation in the hope that providing an affordance for the user to execute a feedback action will lead to better long term recommendations. However, with the explanation-action tradeoff, even if the system is fortunate when taking a chance by providing a less faithful explanation that successfully solicits user feedback, the system will \textit{never} want to repeat the specific explanation-action in the future. We therefore highlight the explanation-action tradeoff as an important consideration when implementing an advice-taking system.

While the explanation-action tradeoff was observed in our second case study, it is important to clarify why we did not observe a similar tradeoff in our first case study in the image domain. We believe this to be a consequence of fundamental differences between our case studies, as detailed in Table \ref{tab:casestudies}. First, the explanatory vocabulary is not fully \textit{discoverable} with papers (we only surface 3 terms out of thousands), but it is fully discoverable with images (all constituent superpixels can easily be viewed simultaneously). Second, advice taking in Semantic Sanity was \textit{iterative} because users repeatedly refined their feeds, whereas only one round of advice taking was performed in our image domain experiment, providing no chance for a feedback cycle to develop. Lastly, the explanatory vocabulary is \textit{shared} across papers, but not across images: when providing advice multiple times in the image domain, a different instance will be surfaced each time, meaning that a new set of superpixel features are available for eliciting advice. Given these differences, we would not expect to observe the explanation-action tradeoff in the image domain. However, it is evident that the explanation-action tradeoff may arise in enough advice-taking settings that documenting and investigating it is an important contribution.

\subsection{Decoupling the Effect of Explanations \& Advice Taking}
Previous studies have shown that users prefer recommendations with explanations over recommendations alone \cite{tintarev_masthoff_07, zhang_explainable_2018}. In our user study, we did not include an ``explanations only'' baseline, which would have helped to isolate the contribution of explanations in the preference for our \sys\ system among participants. However, we did analyze the user study results post-hoc to investigate this question. In particular, we studied the results in Tables \ref{tab:ratingsone} and \ref{tab:ratingstwo} in order to assess whether participants' self-reported preferences for our \sys\ system over the baseline system correlated with utilization of the \sys\ affordance for rating topics. The participants who voted \sys\ higher on trust, transparency, intuitiveness, and confidence in not missing papers performed $5.4$, $4.6$, $-0.5$, and $3.8$ more topic annotations, on average, than those who voted the baseline higher, respectively. This suggests that the positive outcomes for those metrics were {\em not} a result of the explanations alone, but were influenced by the advice-taking affordance of \sys.

In Figure \ref{fig:topic_ratings}, we investigate how the number of topic ratings used by each participant varies as a function of their Likert scale ratings in Table \ref{tab:ratingstwo}. We find that a higher overall rating of our \sys\ system and a higher self-reported likelihood of using our \sys\ system in the future are correlated with using more topic annotations (i.e., giving more advice). This indicates that more usage of the \sys\ affordance correlates with a more positive perception of the \sys\ system.

\section{Related Work}
\label{sec:relatedwork}

Space precludes a discussion of work on explanation generation; we focus our description of prior work on approaches for incorporating human advice in machine learning models and on approaches for creating pseudo-instances by labeling features. Some work transcends these distinctions, however; Smith-Renner \etal~\cite{SmithRenner2020NoEW} show that many users expect that an ML model will improve over time, and that users are frustrated with imperfect AI systems that provide explanations without supporting the ability to receive corresponding feedback. Furthermore, there are other general-purpose ways to improve model accuracy besides high-level advice: labeling new training instances, altering the weights of training instances, and providing an `undo' button to remove a label that has just been added are a few such methods.

\subsection{Enabling  Machine Learners to Take Human Advice: Interpretable Models}

Research from interactive machine learning and human-AI interaction has shown the benefits of enabling learning models, including both recommender systems and classifiers, to take advice from humans \cite{amershi_2012,simard_machine_2017}. For example, Lou~\textit{et~al.} ~\shortcite{caruana_2}, Lou~\textit{et~al.} ~\shortcite{caruana_1}, Caruana~\textit{et~al.} ~\shortcite{caruana_3}, and Wang~\textit{et~al.}~\shortcite{wang2021} have demonstrated the value of GAMs and GA$^2$Ms, which can be directly modified by humans via the alteration of shape functions. Likewise, Kulesza~\textit{et~al.}~\shortcite{kulesza_principles_2015} have shown the power of explanatory debugging of models. However, this research has focused on transparent, interpretable models, where the models can be adjusted directly \cite{weld_challenge_2018}. \sys\ extends the paradigm of interactive machine learning and advice taking to opaque models. Moreover, these evaluations often focus on user ratings rather than quantitative demonstrations of a model's improvement in accuracy via advice taking. As argued by \cite{teso_2022_survey}, benchmarking and evaluation remain open problems in leveraging explanations in interactive machine learning, one form of advice taking. In our second case study, we directly quantitatively evaluate \sys\ accuracy improvements relative to a baseline.

Recommender systems are a common domain for studying explainability and advice taking due to the feedback loop and interactivity essential to the task of recommendation~\cite{ahn_2015,brusilovsky_2020,he_interactive_2016,loepp_interactive_2016,pu_2011,tintarev_designing_2011,tsai_2018,tsai_social_2019,waern_user_2004,zhang_explainable_2018}. Some recommender systems take a human's advice via affordances other than rating content \cite{he_interactive_2016}. The majority of these systems enable advice taking in response to a global explanation of the system's behavior \cite{bakalov_2013, bostandjiev_tasteweights_2012, bostandjiev_linkedvis_2013, bruns_what_2015, gretarsson_2010, jin_effects_2018, kangasraasio_2015, rosenthal_2010, schaffer_2015, odonovan_2008, parra_user-controllable_2015, knijnenburg_2011, tsai_2020}. Others enable advice taking in response to instance-level explanations or no explanations at all \cite{ahn_2007, harper_putting_2015, kulesza_tell_2012, vig_tag_2012}. The combined affordances of advice taking and explainability can lead to a higher degree of user satisfaction \cite{herlocker_explaining_2000}; more trust in and perceived control of the system \cite{cramer_effects_2008,herlocker_explaining_2000,pu_2006,verbert_2013}; and better mental models, without significantly increasing the cognitive load \cite{kulezsa_too_2013,kulesza_principles_2015,rosenthal_2010}. In contrast to LIMEADE, however, all of this work either relies on interpretable models or implements advice taking in an algorithmic-specific fashion that is not extensible to an arbitrary opaque machine-learned model. 

\subsection{Enabling Machine Learners to Take Human Advice: Architecture-Specific Models}

Other work has explored the extension of advice taking to specific classes of opaque models, such as neural architectures. Like LIMEADE, the methods proposed by both Rieger  \etal~\cite{rieger} and Ross \etal~\cite{doshivelez} accept human input in response to advice given in terms of an explanatory vocabulary, but their methods are restricted to differentiable models whose gradients can be accessed. 
Rieger  \etal\ modify the loss function in order to incorporate a ``contextual decomposition explanation penalization'' that encodes a human's domain knowledge in response to an explanation; and Ross \etal\ modify the loss function through input gradient penalization as a form of regularization. 
However, both methods are largely evaluated with simulated experiments on small, artificial datasets, where the confounds are often synthetically generated. With DECOY-MNIST, for example, the training data is artificially colored systematically, leading a learner to recognize color rather than shapes. Some methods can effectively adjust the loss function to reflect advice like ``ignore color'' yielding more robust behavior, but in the real world confounds are much more complex, and it is not clear that these methods generalize well, even for their specific architecture classes. 

Liu and Avci \cite{liu} present an NLP-specific method that allows a developer to introduce a term into the loss function that can counteract biases exposed by explanations. Specifically, the method can be used to guide a hate-speech detection model away from overly relying on tokens (such as `gay') associated with protected groups. This is different from the feedback accepted by \sys, since it says ``Ignore this feature,'' rather than ``Consider this feature to be positive/negative,'' but it is an important type of high-level advice. Liu and Avci tested their approach on both a synthetic and  real-world domain, showing modest improvement on the latter. Unlike LIMEADE, however, their approach works only for neural models and has only been tested on an NLP toxicity detection task.

In computer vision models, researchers have created methods for analyzing the behavior of specific neurons, \eg\ discovering one that produces foliage in a generative model; follow-on research has developed methods for similarly editing these models by rewriting the behavior of those neurons~\cite{bau2020rewriting,mu_2020}. While impressive, these models are both domain and architecture specific, and require great expertise on the part of the user --- far from McCarthy's dream.
 
\subsection{Enabling Machine Learners to Take Human Advice: Arbitrary Opaque Models}
\label{sec:opaque-advice}

Dasgupta \etal~\cite{teaching_blackbox_learner} consider the problem of teaching an opaque learner whose representation and hypothesis class are unknown. The authors show that by interacting with the black-box learner, a teacher can efficiently find a good set of teaching instances. However, Dasgupta \etal's approach is highly theoretical and assumes a noiseless version-space formulation of learning, where the concept is perfectly learnable. Most importantly, in contrast to \sys, their method doesn't enable the teacher to provide advice in a high-level language.

Broadly speaking, the advice-taking interaction in LIMEADE is similar to classical human-in-the-loop active learning (AL) \cite{settles2009active}, which includes techniques that are applicable to opaque models.  However, LIMEADE is distinct from typical AL in that the user is not limited to labeling instances, but can give advice on how the interpretable features should be driving model behavior (which are converted into pseudo-labeled instances using our approach).  Further, AL work focuses on algorithms to select informative instances for labeling, whereas \sys\ creates affordances for feedback on top of explanations that {\em the user} may choose to act upon.

Closest to our work, Schramowski \etal~\cite{schramowski} present a method for adding a user into the ML training loop in order to see the AI’s explanations and provide feedback to improve decision making. Like \sys, their method works with an arbitrary opaque classifier, requiring only the ability to add new instances to the training set.  Furthermore, they also interpret human feedback in the vocabulary used in an arbitrary, explanatory model, such as that produced by LIME~\cite{Ribeiro_LIME_2016}.
However, unlike our work, Schramowski \etal\ do not provide a way for the human to explain to the AI why it made a mistake. Instead, they focus on corrections for when the model is ``right for the wrong reason.'' Like \sys, their method generates pseudo-instances, called ``counter-examples,'' that are created by altering the selected feature of the explained instance in order to reduce confounds (including through randomization, a change to an alternative value, or a substitution with the value for that component appearing in other training instances of the same class). 
Furthermore, Schramowski \etal\ include only a single experiment to demonstrate their model-agnostic method: on a version of the toy MNIST dataset that was artificially biased to include decoy pixels (Table 1a ~\cite{schramowski}); their other experiments used a version of Ross \etal's neural-specific loss~\cite{doshivelez}.

\subsection{Labeling Features and Creating Pseudo-instances} 

While canonical methods of feedback involve providing additional labeled instances \cite{wallace_2019_trick}, one approach to semi-supervised learning involves training a machine learner on labeled instances as well as labeled \textit{features}~\cite{wu_04, Schapire2005BoostingWP, druck_08, raghavan_07, godbole_04, liu_04}. In the text classification setting, this often takes the form of labeling $n$-gram features. These features are then used to construct pseudo-instances (\eg,  documents containing just the labeled $n$-gram itself, labeled according to the feature's assigned label) or to power methods such as the generalized expectation criteria~\cite{wu_04}.  \sys\ extends this semi-supervised approach by translating feature labels in an explanatory model into pseudo-instances for retraining a much more complex  opaque model, which is represented using different features.
\section{Conclusion \& A Call to Action}\label{sec:conclusion}

To be effective partners in a human-AI team, an AI system must be able to not only explain its decisions but also take advice given by humans in terms of that explanation. While interpretable classifiers such as GAMs support explanation-based advice taking, and post-hoc methods such as LIME provide {\em explanations} for opaque ML models, we present the first method for {\em updating} an arbitrary opaque model using positive and negative advice given in terms of a high-level vocabulary (such as the featurization of an explanatory model). Furthermore, we are the first to evaluate such a method on a large number (70) of real-world domains and with user studies. In our first case study, we  used \sys\ to implement advice taking on twenty image classification domains. We showed significant improvement over a strong baseline in the two-shot case. In our second case study, we incorporated \sys\ into  \sanity, a publicly-available computer science research paper recommender.  Significantly, this case study adopted a different domain and different task, demonstrating how \sys\ is a general framework for advice taking. Our user study over 21 participants demonstrated that users strongly prefer our advice-taking system, lauding perceived control and their sense of trust. While we failed to show improved accuracy of the resulting recommender for these users, as measured with DCG, a study of the long-term logs of 30 different, organic users did show significantly improved NDCG. Furthermore, another log study uncovered a fundamental tension between canonical explanation approaches that greedily select the most influential features and those that provide the best affordances for advice taking.

Much work remains to be done. We hope to develop improved methods for interpreting human advice and better understand when such advice is useful. Experiments using different explanatory vocabularies would also be useful. Additional questions, such as simultaneous advice taking from multiple people in the non-personalized setting, are worth pursuing. Furthermore, developing other forms of advice taking remains a fruitful area for exploration. For example, enabling humans to give advice by adding features, or communicating through natural language or other forms of communication, are understudied challenges. Moreover, understanding  various ``hyperparameters'' surrounding advice taking, such as the proper strength of an update, remains an important question both empirically and theoretically. For example, in the case of recommenders, should the strength of an advice-taking update be personalized? Should it change during different stages of updating a model? Are other methods effective at combating the explanation-action tradeoff, such as using arbitrary English feedback to generate a pseudo-instance, rather than restricting to advice written using the features surfaced in greedy explanations? While we did not evaluate \sys\ according to improvements in fairness, robustness, or model compliance, advice taking could be used for these purposes, and another compelling direction of research concerns refining and evaluating advice-taking frameworks in this context. Lastly, it is important to further investigate the entanglement between training data and supervision with respect to advice taking, as described in Section \ref{sec:entanglement}.

We consider our paper a ``Call to Action'' for researchers in human-AI interaction to study the advice-taking problem for opaque machine learners. From search \& recommendation to image recognition to medical diagnosis, opaque machine learners are ubiquitous. End-users deserve new methods for adjusting these machine learning systems by giving advice in terms of a high-level vocabulary. To aid future research, we will release the code written for our image domain experiments, including our our modified COCO dataset with the precomputed superpixel vocabulary and corresponding embeddings, as well as our functioning implementation of \sys. We hope that this work will contribute to opening a new direction of research in human-AI interaction devoted to this challenging and pressing problem.

\begin{acks}
The authors would like to thank Sam Skjonsberg and Daniel King for their help in setting up the initial Semantic Sanity prototype; Matt Latzke for his interface design work; Chelsea Haupt and Sebastian Kohlmeier for their management of platform  development; Alex Schokking for his work on scaling Semantic Sanity for public launch; Arman Cohan and Sergey Feldman for providing the neural paper embeddings; Yogi Chandrasekhar and Chris Wilhelm for their guidance with implementing \sys\ for Semantic Sanity; Chris Wilhelm for his help in implementing functionality necessary for the user study; Ani Kembhavi for his advice regarding \sys\ in the image domain; and Iz Beltagy, Jonathan Bragg,  Krzysztof Gajos, Rao Kambhampati, and Marco Tulio Ribeiro for their helpful feedback. We would also like to thank Andrej Karpathy for his arXiv Sanity Preserver platform, which served as the basis for Semantic Sanity. Lastly, we thank the many users of Semantic Sanity for their feedback. This material is based upon work supported by the National Science Foundation Graduate Research Fellowship under Grant DGE-1762114, by ONR grant {N00014-18-1-2193}, NSF RAPID grant 2040196, the WRF/Cable Professorship, and the Allen Institute for Artificial Intelligence (AI2).
\end{acks}

\bibliographystyle{ACM-Reference-Format}
\bibliography{references}


\begin{thebibliography}{91}


\ifx \showCODEN    \undefined \def \showCODEN     #1{\unskip}     \fi
\ifx \showDOI      \undefined \def \showDOI       #1{#1}\fi
\ifx \showISBNx    \undefined \def \showISBNx     #1{\unskip}     \fi
\ifx \showISBNxiii \undefined \def \showISBNxiii  #1{\unskip}     \fi
\ifx \showISSN     \undefined \def \showISSN      #1{\unskip}     \fi
\ifx \showLCCN     \undefined \def \showLCCN      #1{\unskip}     \fi
\ifx \shownote     \undefined \def \shownote      #1{#1}          \fi
\ifx \showarticletitle \undefined \def \showarticletitle #1{#1}   \fi
\ifx \showURL      \undefined \def \showURL       {\relax}        \fi
\providecommand\bibfield[2]{#2}
\providecommand\bibinfo[2]{#2}
\providecommand\natexlab[1]{#1}
\providecommand\showeprint[2][]{arXiv:#2}

\bibitem[\protect\citeauthoryear{Ahn, Brusilovsky, Grady, He, and Syn}{Ahn
  et~al\mbox{.}}{2007}]%
        {ahn_2007}
\bibfield{author}{\bibinfo{person}{Jae-wook Ahn}, \bibinfo{person}{Peter
  Brusilovsky}, \bibinfo{person}{Jonathan Grady}, \bibinfo{person}{Daqing He},
  {and} \bibinfo{person}{Sue~Yeon Syn}.} \bibinfo{year}{2007}\natexlab{}.
\newblock \showarticletitle{Open User Profiles for Adaptive News Systems: Help
  or Harm?}. In \bibinfo{booktitle}{\emph{WWW '07}} (Banff, Alberta, Canada).
  \bibinfo{publisher}{ACM}, \bibinfo{pages}{11--20}.
\newblock
\showISBNx{978-1-59593-654-7}
\urldef\tempurl%
\url{https://doi.org/10.1145/1242572.1242575}
\showDOI{\tempurl}


\bibitem[\protect\citeauthoryear{Ahn, Brusilovsky, and Han}{Ahn
  et~al\mbox{.}}{2015}]%
        {ahn_2015}
\bibfield{author}{\bibinfo{person}{Jae-wook Ahn}, \bibinfo{person}{Peter
  Brusilovsky}, {and} \bibinfo{person}{Shuguang Han}.}
  \bibinfo{year}{2015}\natexlab{}.
\newblock \showarticletitle{Personalized Search: Reconsidering the Value of
  Open User Models}. In \bibinfo{booktitle}{\emph{IUI '15}} (Atlanta, USA).
  \bibinfo{publisher}{ACM}, \bibinfo{pages}{202--212}.
\newblock
\showISBNx{978-1-4503-3306-1}
\urldef\tempurl%
\url{https://doi.org/10.1145/2678025.2701410}
\showDOI{\tempurl}


\bibitem[\protect\citeauthoryear{Akata, Reed, Walter, Lee, and Schiele}{Akata
  et~al\mbox{.}}{2015}]%
        {Akata2015EvaluationOO}
\bibfield{author}{\bibinfo{person}{Zeynep Akata}, \bibinfo{person}{Scott~E.
  Reed}, \bibinfo{person}{Daniel Walter}, \bibinfo{person}{Honglak Lee}, {and}
  \bibinfo{person}{Bernt Schiele}.} \bibinfo{year}{2015}\natexlab{}.
\newblock \showarticletitle{Evaluation of output embeddings for fine-grained
  image classification}.
\newblock \bibinfo{journal}{\emph{2015 IEEE Conference on Computer Vision and
  Pattern Recognition (CVPR)}} (\bibinfo{year}{2015}),
  \bibinfo{pages}{2927--2936}.
\newblock


\bibitem[\protect\citeauthoryear{Amershi, Cakmak, Knox, and Kulesza}{Amershi
  et~al\mbox{.}}{2014}]%
        {amershi_power_2014}
\bibfield{author}{\bibinfo{person}{Saleema Amershi}, \bibinfo{person}{Maya
  Cakmak}, \bibinfo{person}{William~Bradley Knox}, {and} \bibinfo{person}{Todd
  Kulesza}.} \bibinfo{year}{2014}\natexlab{}.
\newblock \showarticletitle{Power to the {People}: {The} {Role} of {Humans} in
  {Interactive} {Machine} {Learning}}.
\newblock \bibinfo{journal}{\emph{AI Magazine}} \bibinfo{volume}{35},
  \bibinfo{number}{4} (\bibinfo{year}{2014}), \bibinfo{pages}{105--120}.
\newblock
\urldef\tempurl%
\url{https://doi.org/10.1609/aimag.v35i4.2513}
\showDOI{\tempurl}


\bibitem[\protect\citeauthoryear{Amershi, Fogarty, and Weld}{Amershi
  et~al\mbox{.}}{2012}]%
        {amershi_2012}
\bibfield{author}{\bibinfo{person}{Saleema Amershi}, \bibinfo{person}{James
  Fogarty}, {and} \bibinfo{person}{Daniel Weld}.}
  \bibinfo{year}{2012}\natexlab{}.
\newblock \showarticletitle{Regroup: Interactive Machine Learning for On-demand
  Group Creation in Social Networks}. In \bibinfo{booktitle}{\emph{CHI '12}}
  (Austin, USA). \bibinfo{publisher}{ACM}, \bibinfo{pages}{21--30}.
\newblock
\showISBNx{978-1-4503-1015-4}
\urldef\tempurl%
\url{https://doi.org/10.1145/2207676.2207680}
\showDOI{\tempurl}


\bibitem[\protect\citeauthoryear{Amershi, Weld, Vorvoreanu, Fourney, Nushi,
  Collisson, Suh, Iqbal, Bennett, Inkpen, and et~al.}{Amershi
  et~al\mbox{.}}{2019}]%
        {Amershi2019GuidelinesFH}
\bibfield{author}{\bibinfo{person}{Saleema Amershi}, \bibinfo{person}{Dan
  Weld}, \bibinfo{person}{Mihaela Vorvoreanu}, \bibinfo{person}{Adam Fourney},
  \bibinfo{person}{Besmira Nushi}, \bibinfo{person}{Penny Collisson},
  \bibinfo{person}{Jina Suh}, \bibinfo{person}{Shamsi Iqbal},
  \bibinfo{person}{Paul Bennett}, \bibinfo{person}{Kori Inkpen}, {and}
  \bibinfo{person}{et al.}} \bibinfo{year}{2019}\natexlab{}.
\newblock \showarticletitle{Guidelines for Human-AI Interaction}. In
  \bibinfo{booktitle}{\emph{CHI '19}} (Glasgow, Scotland).
  \bibinfo{publisher}{ACM}, Article \bibinfo{articleno}{3}.
\newblock
\showISBNx{9781450359702}
\urldef\tempurl%
\url{https://doi.org/10.1145/3290605.3300233}
\showDOI{\tempurl}


\bibitem[\protect\citeauthoryear{Ammar, Groeneveld, Bhagavatula, Beltagy,
  Crawford, Downey, Dunkelberger, and et~al.}{Ammar et~al\mbox{.}}{2018}]%
        {Ammar2018ConstructionOT}
\bibfield{author}{\bibinfo{person}{Waleed Ammar}, \bibinfo{person}{Dirk
  Groeneveld}, \bibinfo{person}{Chandra Bhagavatula}, \bibinfo{person}{Iz
  Beltagy}, \bibinfo{person}{Miles Crawford}, \bibinfo{person}{Doug Downey},
  \bibinfo{person}{Jason Dunkelberger}, {and} \bibinfo{person}{et al.}}
  \bibinfo{year}{2018}\natexlab{}.
\newblock \showarticletitle{Construction of the Literature Graph in Semantic
  Scholar}. In \bibinfo{booktitle}{\emph{NAACL-HLT '18}} (New Orleans, USA).
  \bibinfo{publisher}{ACL}, \bibinfo{pages}{84--91}.
\newblock
\urldef\tempurl%
\url{https://doi.org/10.18653/v1/N18-3011}
\showDOI{\tempurl}


\bibitem[\protect\citeauthoryear{Bakalov, Meurs, K\"{o}nig-Ries, Sateli, Witte,
  Butler, and Tsang}{Bakalov et~al\mbox{.}}{2013}]%
        {bakalov_2013}
\bibfield{author}{\bibinfo{person}{Fedor Bakalov}, \bibinfo{person}{Marie-Jean
  Meurs}, \bibinfo{person}{Birgitta K\"{o}nig-Ries}, \bibinfo{person}{Bahar
  Sateli}, \bibinfo{person}{Ren{\'e} Witte}, \bibinfo{person}{Greg Butler},
  {and} \bibinfo{person}{Adrian Tsang}.} \bibinfo{year}{2013}\natexlab{}.
\newblock \showarticletitle{An Approach to Controlling User Models and
  Personalization Effects in Recommender Systems}. In
  \bibinfo{booktitle}{\emph{IUI '13}} (Santa Monica, CA).
  \bibinfo{publisher}{ACM}, \bibinfo{pages}{49--56}.
\newblock
\showISBNx{978-1-4503-1965-2}
\urldef\tempurl%
\url{https://doi.org/10.1145/2449396.2449405}
\showDOI{\tempurl}


\bibitem[\protect\citeauthoryear{Bau, Liu, Wang, Zhu, and Torralba}{Bau
  et~al\mbox{.}}{2020}]%
        {bau2020rewriting}
\bibfield{author}{\bibinfo{person}{David Bau}, \bibinfo{person}{Steven Liu},
  \bibinfo{person}{Tongzhou Wang}, \bibinfo{person}{Jun-Yan Zhu}, {and}
  \bibinfo{person}{Antonio Torralba}.} \bibinfo{year}{2020}\natexlab{}.
\newblock \showarticletitle{Rewriting a Deep Generative Model}. In
  \bibinfo{booktitle}{\emph{Proceedings of the European Conference on Computer
  Vision (ECCV)}}.
\newblock


\bibitem[\protect\citeauthoryear{Beel, Gipp, Langer, and Breitinger}{Beel
  et~al\mbox{.}}{2016}]%
        {beel_research-paper_2016}
\bibfield{author}{\bibinfo{person}{Joeran Beel}, \bibinfo{person}{Bela Gipp},
  \bibinfo{person}{Stefan Langer}, {and} \bibinfo{person}{Corinna Breitinger}.}
  \bibinfo{year}{2016}\natexlab{}.
\newblock \showarticletitle{Research-paper recommender systems: a literature
  survey}.
\newblock \bibinfo{journal}{\emph{International Journal on Digital Libraries}}
  \bibinfo{volume}{17}, \bibinfo{number}{4} (\bibinfo{year}{2016}),
  \bibinfo{pages}{305--338}.
\newblock
\showISSN{1432-1300}
\urldef\tempurl%
\url{https://doi.org/10.1007/s00799-015-0156-0}
\showDOI{\tempurl}


\bibitem[\protect\citeauthoryear{Beltagy, Lo, and Cohan}{Beltagy
  et~al\mbox{.}}{2019}]%
        {beltagy-etal-2019-scibert}
\bibfield{author}{\bibinfo{person}{Iz Beltagy}, \bibinfo{person}{Kyle Lo},
  {and} \bibinfo{person}{Arman Cohan}.} \bibinfo{year}{2019}\natexlab{}.
\newblock \showarticletitle{SciBERT: A Pretrained Language Model for Scientific
  Text}. In \bibinfo{booktitle}{\emph{EMNLP '19}}.
\newblock


\bibitem[\protect\citeauthoryear{Bhagavatula, Feldman, Power, and
  Ammar}{Bhagavatula et~al\mbox{.}}{2018}]%
        {bhagavatula-etal-2018-content}
\bibfield{author}{\bibinfo{person}{Chandra Bhagavatula},
  \bibinfo{person}{Sergey Feldman}, \bibinfo{person}{Russell Power}, {and}
  \bibinfo{person}{Waleed Ammar}.} \bibinfo{year}{2018}\natexlab{}.
\newblock \showarticletitle{Content-Based Citation Recommendation}. In
  \bibinfo{booktitle}{\emph{ACL '18}}. \bibinfo{publisher}{ACL},
  \bibinfo{address}{New Orleans, LA}, \bibinfo{pages}{238--251}.
\newblock
\urldef\tempurl%
\url{https://doi.org/10.18653/v1/N18-1022}
\showDOI{\tempurl}


\bibitem[\protect\citeauthoryear{Bostandjiev, O'Donovan, and
  H\"{o}llerer}{Bostandjiev et~al\mbox{.}}{2012}]%
        {bostandjiev_tasteweights_2012}
\bibfield{author}{\bibinfo{person}{Svetlin Bostandjiev}, \bibinfo{person}{John
  O'Donovan}, {and} \bibinfo{person}{Tobias H\"{o}llerer}.}
  \bibinfo{year}{2012}\natexlab{}.
\newblock \showarticletitle{TasteWeights: A Visual Interactive Hybrid
  Recommender System}. In \bibinfo{booktitle}{\emph{RecSys '12}} (Dublin,
  Ireland). \bibinfo{publisher}{ACM}, \bibinfo{pages}{35--42}.
\newblock
\showISBNx{978-1-4503-1270-7}
\urldef\tempurl%
\url{https://doi.org/10.1145/2365952.2365964}
\showDOI{\tempurl}


\bibitem[\protect\citeauthoryear{Bostandjiev, O'Donovan, and
  H\"ollerer}{Bostandjiev et~al\mbox{.}}{2013}]%
        {bostandjiev_linkedvis_2013}
\bibfield{author}{\bibinfo{person}{Svetlin Bostandjiev}, \bibinfo{person}{John
  O'Donovan}, {and} \bibinfo{person}{Tobias H\"ollerer}.}
  \bibinfo{year}{2013}\natexlab{}.
\newblock \showarticletitle{{LinkedVis}: exploring social and semantic career
  recommendations}. In \bibinfo{booktitle}{\emph{{IUI} '13}} (Santa Monica,
  USA). \bibinfo{publisher}{ACM}, \bibinfo{pages}{107}.
\newblock
\showISBNx{978-1-4503-1965-2}
\urldef\tempurl%
\url{https://doi.org/10.1145/2449396.2449412}
\showDOI{\tempurl}


\bibitem[\protect\citeauthoryear{Bruns, Valdez, Greven, Ziefle, and
  Schroeder}{Bruns et~al\mbox{.}}{2015}]%
        {bruns_what_2015}
\bibfield{author}{\bibinfo{person}{Simon Bruns},
  \bibinfo{person}{Andr\'e~Calero Valdez}, \bibinfo{person}{Christoph Greven},
  \bibinfo{person}{Martina Ziefle}, {and} \bibinfo{person}{Ulrik Schroeder}.}
  \bibinfo{year}{2015}\natexlab{}.
\newblock \showarticletitle{What {Should} {I} {Read} {Next}? {A} {Personalized}
  {Visual} {Publication} {Recommender} {System}}. In
  \bibinfo{booktitle}{\emph{Human {Interface} and the {Management} of
  {Information}. {Information} and {Knowledge} in {Context}}}.
  \bibinfo{publisher}{Springer}, \bibinfo{pages}{89--100}.
\newblock
\showISBNx{978-3-319-20618-9}


\bibitem[\protect\citeauthoryear{Brusilovsky, de~Gemmis, Felfernig, Lops,
  O'Donovan, Semeraro, and Willemsen}{Brusilovsky et~al\mbox{.}}{2020}]%
        {brusilovsky_2020}
\bibfield{author}{\bibinfo{person}{Peter Brusilovsky}, \bibinfo{person}{Marco
  de Gemmis}, \bibinfo{person}{Alexander Felfernig}, \bibinfo{person}{Pasquale
  Lops}, \bibinfo{person}{John O'Donovan}, \bibinfo{person}{Giovanni Semeraro},
  {and} \bibinfo{person}{Martijn~C. Willemsen}.}
  \bibinfo{year}{2020}\natexlab{}.
\newblock \showarticletitle{Interfaces and Human Decision Making for
  Recommender Systems}. In \bibinfo{booktitle}{\emph{RecSys '20}} (Virtual
  Event, Brazil). \bibinfo{pages}{613–618}.
\newblock
\showISBNx{9781450375832}
\urldef\tempurl%
\url{https://doi.org/10.1145/3383313.3411539}
\showURL{%
\tempurl}


\bibitem[\protect\citeauthoryear{Caruana, Lou, Gehrke, Koch, Sturm, and
  Elhadad}{Caruana et~al\mbox{.}}{2015}]%
        {caruana_3}
\bibfield{author}{\bibinfo{person}{Rich Caruana}, \bibinfo{person}{Yin Lou},
  \bibinfo{person}{Johannes Gehrke}, \bibinfo{person}{Paul Koch},
  \bibinfo{person}{Marc Sturm}, {and} \bibinfo{person}{Noemie Elhadad}.}
  \bibinfo{year}{2015}\natexlab{}.
\newblock \showarticletitle{Intelligible Models for HealthCare: Predicting
  Pneumonia Risk and Hospital 30-Day Readmission}. In
  \bibinfo{booktitle}{\emph{KDD '15}} (Sydney, NSW, Australia).
  \bibinfo{publisher}{ACM}, \bibinfo{pages}{1721–1730}.
\newblock
\showISBNx{9781450336642}
\urldef\tempurl%
\url{https://doi.org/10.1145/2783258.2788613}
\showDOI{\tempurl}


\bibitem[\protect\citeauthoryear{Cohan, Feldman, Beltagy, Downey, and
  Weld}{Cohan et~al\mbox{.}}{2020a}]%
        {cohan2020}
\bibfield{author}{\bibinfo{person}{Arman Cohan}, \bibinfo{person}{Sergey
  Feldman}, \bibinfo{person}{Iz Beltagy}, \bibinfo{person}{Doug Downey}, {and}
  \bibinfo{person}{Daniel Weld}.} \bibinfo{year}{2020}\natexlab{a}.
\newblock \bibinfo{title}{SPECTER: Document-level Representation Learning using
  Citation-informed Transformers}.
\newblock
\newblock
\showeprint[arxiv]{2004.07180}~[cs.CL]


\bibitem[\protect\citeauthoryear{Cohan, Feldman, Beltagy, Downey, and
  Weld}{Cohan et~al\mbox{.}}{2020b}]%
        {specter2020cohan}
\bibfield{author}{\bibinfo{person}{Arman Cohan}, \bibinfo{person}{Sergey
  Feldman}, \bibinfo{person}{Iz Beltagy}, \bibinfo{person}{Doug Downey}, {and}
  \bibinfo{person}{Daniel~S. Weld}.} \bibinfo{year}{2020}\natexlab{b}.
\newblock \showarticletitle{{SPECTER: Document-level Representation Learning
  using Citation-informed Transformers}}. In \bibinfo{booktitle}{\emph{ACL}}.
\newblock


\bibitem[\protect\citeauthoryear{Cramer, Evers, Ramlal, van Someren, Rutledge,
  Stash, Aroyo, and Wielinga}{Cramer et~al\mbox{.}}{2008}]%
        {cramer_effects_2008}
\bibfield{author}{\bibinfo{person}{Henriette Cramer}, \bibinfo{person}{Vanessa
  Evers}, \bibinfo{person}{Satyan Ramlal}, \bibinfo{person}{Maarten van
  Someren}, \bibinfo{person}{Lloyd Rutledge}, \bibinfo{person}{Natalia Stash},
  \bibinfo{person}{Lora Aroyo}, {and} \bibinfo{person}{Bob Wielinga}.}
  \bibinfo{year}{2008}\natexlab{}.
\newblock \showarticletitle{The effects of transparency on trust in and
  acceptance of a content-based art recommender}.
\newblock \bibinfo{journal}{\emph{UMUAI}} \bibinfo{volume}{18},
  \bibinfo{number}{5} (\bibinfo{year}{2008}), \bibinfo{pages}{455}.
\newblock
\showISSN{1573-1391}
\urldef\tempurl%
\url{https://doi.org/10.1007/s11257-008-9051-3}
\showDOI{\tempurl}


\bibitem[\protect\citeauthoryear{Das, Moore, Wong, Stumpf, Oberst, Mcintosh,
  and Burnett}{Das et~al\mbox{.}}{2013}]%
        {das_2013}
\bibfield{author}{\bibinfo{person}{Shubhomoy Das}, \bibinfo{person}{Travis
  Moore}, \bibinfo{person}{Weng-Keen Wong}, \bibinfo{person}{Simone Stumpf},
  \bibinfo{person}{Ian Oberst}, \bibinfo{person}{Kevin Mcintosh}, {and}
  \bibinfo{person}{Margaret Burnett}.} \bibinfo{year}{2013}\natexlab{}.
\newblock \showarticletitle{End-User Feature Labeling: Supervised and
  Semi-Supervised Approaches Based on Locally-Weighted Logistic Regression}.
\newblock \bibinfo{journal}{\emph{Artif. Intell.}}  \bibinfo{volume}{204}
  (\bibinfo{date}{nov} \bibinfo{year}{2013}), \bibinfo{pages}{56–74}.
\newblock
\showISSN{0004-3702}
\urldef\tempurl%
\url{https://doi.org/10.1016/j.artint.2013.08.003}
\showDOI{\tempurl}


\bibitem[\protect\citeauthoryear{Dasgupta, Hsu, Poulis, and Zhu}{Dasgupta
  et~al\mbox{.}}{2019}]%
        {teaching_blackbox_learner}
\bibfield{author}{\bibinfo{person}{Sanjoy Dasgupta}, \bibinfo{person}{Daniel
  Hsu}, \bibinfo{person}{Stefanos Poulis}, {and} \bibinfo{person}{Xiaojin
  Zhu}.} \bibinfo{year}{2019}\natexlab{}.
\newblock \showarticletitle{Teaching a black-box learner}. In
  \bibinfo{booktitle}{\emph{ICML '19}}. \bibinfo{publisher}{PMLR},
  \bibinfo{address}{Long Beach, USA}, \bibinfo{pages}{1547--1555}.
\newblock


\bibitem[\protect\citeauthoryear{Deng, Dong, Socher, Li, Li, and Fei-Fei}{Deng
  et~al\mbox{.}}{2009}]%
        {imagenet}
\bibfield{author}{\bibinfo{person}{J. Deng}, \bibinfo{person}{W. Dong},
  \bibinfo{person}{R. Socher}, \bibinfo{person}{L.-J. Li}, \bibinfo{person}{K.
  Li}, {and} \bibinfo{person}{L. Fei-Fei}.} \bibinfo{year}{2009}\natexlab{}.
\newblock \showarticletitle{{ImageNet: A Large-Scale Hierarchical Image
  Database}}. In \bibinfo{booktitle}{\emph{CVPR09}}.
\newblock


\bibitem[\protect\citeauthoryear{Doshi-Velez and Kim}{Doshi-Velez and
  Kim}{2017}]%
        {doshi-velez_towards_2017}
\bibfield{author}{\bibinfo{person}{Finale Doshi-Velez} {and}
  \bibinfo{person}{Been Kim}.} \bibinfo{year}{2017}\natexlab{}.
\newblock \bibinfo{title}{Towards {A} {Rigorous} {Science} of {Interpretable}
  {Machine} {Learning}}.
\newblock
\newblock
\urldef\tempurl%
\url{http://arxiv.org/abs/1702.08608}
\showURL{%
\tempurl}


\bibitem[\protect\citeauthoryear{Druck, Mann, and McCallum}{Druck
  et~al\mbox{.}}{2008}]%
        {druck_08}
\bibfield{author}{\bibinfo{person}{Gregory Druck}, \bibinfo{person}{Gideon
  Mann}, {and} \bibinfo{person}{Andrew McCallum}.}
  \bibinfo{year}{2008}\natexlab{}.
\newblock \showarticletitle{Learning from Labeled Features Using Generalized
  Expectation Criteria}. In \bibinfo{booktitle}{\emph{SIGIR '08}} (Singapore,
  Singapore). \bibinfo{pages}{595–602}.
\newblock
\showISBNx{9781605581644}
\urldef\tempurl%
\url{https://dl.acm.org/doi/10.1145/1390334.1390436}
\showURL{%
\tempurl}


\bibitem[\protect\citeauthoryear{Ekstrand, Kannan, Stemper, Butler, Konstan,
  and Riedl}{Ekstrand et~al\mbox{.}}{2010}]%
        {ekstrand_recsys}
\bibfield{author}{\bibinfo{person}{Michael Ekstrand}, \bibinfo{person}{Praveen
  Kannan}, \bibinfo{person}{James Stemper}, \bibinfo{person}{John Butler},
  \bibinfo{person}{Joseph Konstan}, {and} \bibinfo{person}{John Riedl}.}
  \bibinfo{year}{2010}\natexlab{}.
\newblock \showarticletitle{Automatically Building Research Reading Lists}. In
  \bibinfo{booktitle}{\emph{RecSys '10}} (Barcelona, Spain).
  \bibinfo{publisher}{ACM}, \bibinfo{pages}{159–166}.
\newblock
\showISBNx{9781605589060}
\urldef\tempurl%
\url{https://doi.org/10.1145/1864708.1864740}
\showDOI{\tempurl}


\bibitem[\protect\citeauthoryear{Godbole, Harpale, Sarawagi, and
  Chakrabarti}{Godbole et~al\mbox{.}}{2004}]%
        {godbole_04}
\bibfield{author}{\bibinfo{person}{Shantanu Godbole}, \bibinfo{person}{Abhay
  Harpale}, \bibinfo{person}{Sunita Sarawagi}, {and} \bibinfo{person}{Soumen
  Chakrabarti}.} \bibinfo{year}{2004}\natexlab{}.
\newblock \showarticletitle{Document Classification Through Interactive
  Supervision of Document and Term Labels}. In \bibinfo{booktitle}{\emph{PKDD
  '04}}, \bibfield{editor}{\bibinfo{person}{Jean-Fran{\c{c}}ois Boulicaut},
  \bibinfo{person}{Floriana Esposito}, \bibinfo{person}{Fosca Giannotti}, {and}
  \bibinfo{person}{Dino Pedreschi}} (Eds.). \bibinfo{pages}{185--196}.
\newblock
\showISBNx{978-3-540-30116-5}


\bibitem[\protect\citeauthoryear{Gretarsson, O'Donovan, Bostandjiev, Hall, and
  H\"ollerer}{Gretarsson et~al\mbox{.}}{2010}]%
        {gretarsson_2010}
\bibfield{author}{\bibinfo{person}{Brynjar Gretarsson}, \bibinfo{person}{John
  O'Donovan}, \bibinfo{person}{Svetlin Bostandjiev},
  \bibinfo{person}{Christopher Hall}, {and} \bibinfo{person}{Tobias
  H\"ollerer}.} \bibinfo{year}{2010}\natexlab{}.
\newblock \showarticletitle{SmallWorlds: Visualizing Social Recommendations}.
\newblock \bibinfo{journal}{\emph{Computer Graphics Forum}}
  \bibinfo{volume}{29}, \bibinfo{number}{3} (\bibinfo{year}{2010}),
  \bibinfo{pages}{833--842}.
\newblock
\urldef\tempurl%
\url{https://doi.org/10.1111/j.1467-8659.2009.01679.x}
\showDOI{\tempurl}


\bibitem[\protect\citeauthoryear{Guidotti, Monreale, Ruggieri, Turini,
  Giannotti, and Pedreschi}{Guidotti et~al\mbox{.}}{2018}]%
        {survey_explaining_blackbox}
\bibfield{author}{\bibinfo{person}{Riccardo Guidotti}, \bibinfo{person}{Anna
  Monreale}, \bibinfo{person}{Salvatore Ruggieri}, \bibinfo{person}{Franco
  Turini}, \bibinfo{person}{Fosca Giannotti}, {and} \bibinfo{person}{Dino
  Pedreschi}.} \bibinfo{year}{2018}\natexlab{}.
\newblock \showarticletitle{A Survey of Methods for Explaining Black Box
  Models}.
\newblock \bibinfo{journal}{\emph{ACM Comput. Surv.}} \bibinfo{volume}{51},
  \bibinfo{number}{5}, Article \bibinfo{articleno}{93} (\bibinfo{year}{2018}).
\newblock
\showISSN{0360-0300}
\urldef\tempurl%
\url{https://doi.org/10.1145/3236009}
\showDOI{\tempurl}


\bibitem[\protect\citeauthoryear{Harper, Xu, Kaur, Condiff, Chang, and
  Terveen}{Harper et~al\mbox{.}}{2015}]%
        {harper_putting_2015}
\bibfield{author}{\bibinfo{person}{F.~Maxwell Harper}, \bibinfo{person}{Funing
  Xu}, \bibinfo{person}{Harmanpreet Kaur}, \bibinfo{person}{Kyle Condiff},
  \bibinfo{person}{Shuo Chang}, {and} \bibinfo{person}{Loren Terveen}.}
  \bibinfo{year}{2015}\natexlab{}.
\newblock \showarticletitle{Putting {Users} in {Control} of their
  {Recommendations}}. In \bibinfo{booktitle}{\emph{RecSys '15}} (Vienna,
  Austria). \bibinfo{publisher}{ACM}, \bibinfo{pages}{3--10}.
\newblock
\showISBNx{978-1-4503-3692-5}
\urldef\tempurl%
\url{https://doi.org/10.1145/2792838.2800179}
\showDOI{\tempurl}


\bibitem[\protect\citeauthoryear{He, Parra, and Verbert}{He
  et~al\mbox{.}}{2016}]%
        {he_interactive_2016}
\bibfield{author}{\bibinfo{person}{Chen He}, \bibinfo{person}{Denis Parra},
  {and} \bibinfo{person}{Katrien Verbert}.} \bibinfo{year}{2016}\natexlab{}.
\newblock \showarticletitle{Interactive recommender systems: {A} survey of the
  state of the art and future research challenges and opportunities}.
\newblock \bibinfo{journal}{\emph{Expert Systems with Applications}}
  \bibinfo{volume}{56} (\bibinfo{year}{2016}), \bibinfo{pages}{9--27}.
\newblock
\showISSN{09574174}
\urldef\tempurl%
\url{https://doi.org/10.1016/j.eswa.2016.02.013}
\showDOI{\tempurl}


\bibitem[\protect\citeauthoryear{He, Zhang, Ren, and Sun}{He
  et~al\mbox{.}}{2015}]%
        {resnet}
\bibfield{author}{\bibinfo{person}{Kaiming He}, \bibinfo{person}{Xiangyu
  Zhang}, \bibinfo{person}{Shaoqing Ren}, {and} \bibinfo{person}{Jian Sun}.}
  \bibinfo{year}{2015}\natexlab{}.
\newblock \showarticletitle{Deep Residual Learning for Image Recognition}.
\newblock \bibinfo{journal}{\emph{arXiv preprint arXiv:1512.03385}}
  (\bibinfo{year}{2015}).
\newblock


\bibitem[\protect\citeauthoryear{He, Pei, Kifer, Mitra, and Giles}{He
  et~al\mbox{.}}{2010}]%
        {he_2010}
\bibfield{author}{\bibinfo{person}{Qi He}, \bibinfo{person}{Jian Pei},
  \bibinfo{person}{Daniel Kifer}, \bibinfo{person}{Prasenjit Mitra}, {and}
  \bibinfo{person}{Lee Giles}.} \bibinfo{year}{2010}\natexlab{}.
\newblock \showarticletitle{Context-aware Citation Recommendation}. In
  \bibinfo{booktitle}{\emph{WWW '10}} (Raleigh, USA). \bibinfo{publisher}{ACM},
  \bibinfo{pages}{421--430}.
\newblock
\showISBNx{978-1-60558-799-8}
\urldef\tempurl%
\url{https://doi.org/10.1145/1772690.1772734}
\showDOI{\tempurl}


\bibitem[\protect\citeauthoryear{Herlocker, Konstan, and Riedl}{Herlocker
  et~al\mbox{.}}{2000}]%
        {herlocker_explaining_2000}
\bibfield{author}{\bibinfo{person}{Jonathan Herlocker}, \bibinfo{person}{Joseph
  Konstan}, {and} \bibinfo{person}{John Riedl}.}
  \bibinfo{year}{2000}\natexlab{}.
\newblock \showarticletitle{Explaining collaborative filtering
  recommendations}. In \bibinfo{booktitle}{\emph{{CSCW} '00}} (Philadelphia,
  USA). \bibinfo{publisher}{ACM}, \bibinfo{pages}{241--250}.
\newblock
\showISBNx{978-1-58113-222-9}
\urldef\tempurl%
\url{https://doi.org/10.1145/358916.358995}
\showDOI{\tempurl}


\bibitem[\protect\citeauthoryear{Holm}{Holm}{1979}]%
        {holm79}
\bibfield{author}{\bibinfo{person}{Sture Holm}.}
  \bibinfo{year}{1979}\natexlab{}.
\newblock \showarticletitle{A Simple Sequentially Rejective Multiple Test
  Procedure}.
\newblock \bibinfo{journal}{\emph{Scandinavian Journal of Statistics}}
  \bibinfo{volume}{6}, \bibinfo{number}{2} (\bibinfo{year}{1979}),
  \bibinfo{pages}{65--70}.
\newblock
\showISSN{03036898, 14679469}


\bibitem[\protect\citeauthoryear{Jin, Tintarev, and Verbert}{Jin
  et~al\mbox{.}}{2018}]%
        {jin_effects_2018}
\bibfield{author}{\bibinfo{person}{Yucheng Jin}, \bibinfo{person}{Nava
  Tintarev}, {and} \bibinfo{person}{Katrien Verbert}.}
  \bibinfo{year}{2018}\natexlab{}.
\newblock \showarticletitle{Effects of personal characteristics on music
  recommender systems with different levels of controllability}. In
  \bibinfo{booktitle}{\emph{{RecSys} '18}} (Vancouver, Canada).
  \bibinfo{publisher}{ACM}, \bibinfo{pages}{13--21}.
\newblock
\showISBNx{978-1-4503-5901-6}
\urldef\tempurl%
\url{https://doi.org/10.1145/3240323.3240358}
\showDOI{\tempurl}


\bibitem[\protect\citeauthoryear{Kambhampati, Sreedharan, Verma, Zha, and
  Guan}{Kambhampati et~al\mbox{.}}{2022}]%
        {rao_2022}
\bibfield{author}{\bibinfo{person}{Subbarao Kambhampati},
  \bibinfo{person}{Sarath Sreedharan}, \bibinfo{person}{Mudit Verma},
  \bibinfo{person}{Yantian Zha}, {and} \bibinfo{person}{Lin Guan}.}
  \bibinfo{year}{2022}\natexlab{}.
\newblock \showarticletitle{Symbols as a Lingua Franca for Bridging Human-AI
  Chasm for Explainable and Advisable AI Systems}.
\newblock \bibinfo{journal}{\emph{Proceedings of the AAAI Conference on
  Artificial Intelligence}} \bibinfo{volume}{36}, \bibinfo{number}{11}
  (\bibinfo{date}{Jun.} \bibinfo{year}{2022}), \bibinfo{pages}{12262--12267}.
\newblock
\urldef\tempurl%
\url{https://doi.org/10.1609/aaai.v36i11.21488}
\showDOI{\tempurl}


\bibitem[\protect\citeauthoryear{Kanakia, Shen, Eide, and Wang}{Kanakia
  et~al\mbox{.}}{2019}]%
        {Kanakia_MAG_2019}
\bibfield{author}{\bibinfo{person}{Anshul Kanakia}, \bibinfo{person}{Zhihong
  Shen}, \bibinfo{person}{Darrin Eide}, {and} \bibinfo{person}{Kuansan Wang}.}
  \bibinfo{year}{2019}\natexlab{}.
\newblock \showarticletitle{A Scalable Hybrid Research Paper Recommender System
  for Microsoft Academic}. In \bibinfo{booktitle}{\emph{WWW' 19}} (San
  Francisco, USA). \bibinfo{publisher}{ACM}, \bibinfo{pages}{2893--2899}.
\newblock
\showISBNx{978-1-4503-6674-8}
\urldef\tempurl%
\url{https://doi.org/10.1145/3308558.3313700}
\showDOI{\tempurl}


\bibitem[\protect\citeauthoryear{Kangasr\"{a}\"{a}si\"{o}, Glowacka, and
  Kaski}{Kangasr\"{a}\"{a}si\"{o} et~al\mbox{.}}{2015}]%
        {kangasraasio_2015}
\bibfield{author}{\bibinfo{person}{Antti Kangasr\"{a}\"{a}si\"{o}},
  \bibinfo{person}{Dorota Glowacka}, {and} \bibinfo{person}{Samuel Kaski}.}
  \bibinfo{year}{2015}\natexlab{}.
\newblock \showarticletitle{Improving Controllability and Predictability of
  Interactive Recommendation Interfaces for Exploratory Search}. In
  \bibinfo{booktitle}{\emph{IUI '15}} (Atlanta, USA). \bibinfo{publisher}{ACM},
  \bibinfo{pages}{247--251}.
\newblock
\showISBNx{978-1-4503-3306-1}
\urldef\tempurl%
\url{https://doi.org/10.1145/2678025.2701371}
\showDOI{\tempurl}


\bibitem[\protect\citeauthoryear{Karpathy}{Karpathy}{2015}]%
        {karpathy_2015}
\bibfield{author}{\bibinfo{person}{Andrej Karpathy}.}
  \bibinfo{year}{2015}\natexlab{}.
\newblock \bibinfo{title}{Arxiv Sanity Preserver}.
\newblock
\newblock
\urldef\tempurl%
\url{http://www.arxiv-sanity.com/}
\showURL{%
\tempurl}


\bibitem[\protect\citeauthoryear{Knijnenburg, Reijmer, and
  Willemsen}{Knijnenburg et~al\mbox{.}}{2011}]%
        {knijnenburg_2011}
\bibfield{author}{\bibinfo{person}{Bart Knijnenburg}, \bibinfo{person}{Niels
  Reijmer}, {and} \bibinfo{person}{Martijn Willemsen}.}
  \bibinfo{year}{2011}\natexlab{}.
\newblock \showarticletitle{Each to His Own: How Different Users Call for
  Different Interaction Methods in Recommender Systems}. In
  \bibinfo{booktitle}{\emph{RecSys '11}} (Chicago, USA).
  \bibinfo{publisher}{ACM}, \bibinfo{pages}{141--148}.
\newblock
\showISBNx{978-1-4503-0683-6}
\urldef\tempurl%
\url{https://doi.org/10.1145/2043932.2043960}
\showDOI{\tempurl}


\bibitem[\protect\citeauthoryear{Koh, Nguyen, Tang, Mussmann, Pierson, Kim, and
  Liang}{Koh et~al\mbox{.}}{2020}]%
        {Koh2020ConceptBM}
\bibfield{author}{\bibinfo{person}{Pang~Wei Koh}, \bibinfo{person}{Thao
  Nguyen}, \bibinfo{person}{Yew~Siang Tang}, \bibinfo{person}{Stephen
  Mussmann}, \bibinfo{person}{Emma Pierson}, \bibinfo{person}{Been Kim}, {and}
  \bibinfo{person}{Percy Liang}.} \bibinfo{year}{2020}\natexlab{}.
\newblock \showarticletitle{Concept Bottleneck Models}.
\newblock \bibinfo{journal}{\emph{ArXiv}}  \bibinfo{volume}{abs/2007.04612}
  (\bibinfo{year}{2020}).
\newblock


\bibitem[\protect\citeauthoryear{Kulesza, Burnett, Wong, and Stumpf}{Kulesza
  et~al\mbox{.}}{2015}]%
        {kulesza_principles_2015}
\bibfield{author}{\bibinfo{person}{Todd Kulesza}, \bibinfo{person}{Margaret
  Burnett}, \bibinfo{person}{Weng-Keen Wong}, {and} \bibinfo{person}{Simone
  Stumpf}.} \bibinfo{year}{2015}\natexlab{}.
\newblock \showarticletitle{Principles of Explanatory Debugging to Personalize
  Interactive Machine Learning}. In \bibinfo{booktitle}{\emph{IUI '15}}
  (Atlanta, USA). \bibinfo{publisher}{ACM}, \bibinfo{pages}{126--137}.
\newblock
\showISBNx{978-1-4503-3306-1}
\urldef\tempurl%
\url{https://doi.org/10.1145/2678025.2701399}
\showDOI{\tempurl}


\bibitem[\protect\citeauthoryear{Kulesza, Stumpf, Burnett, and Kwan}{Kulesza
  et~al\mbox{.}}{2012}]%
        {kulesza_tell_2012}
\bibfield{author}{\bibinfo{person}{Todd Kulesza}, \bibinfo{person}{Simone
  Stumpf}, \bibinfo{person}{Margaret Burnett}, {and} \bibinfo{person}{Irwin
  Kwan}.} \bibinfo{year}{2012}\natexlab{}.
\newblock \showarticletitle{Tell me more?: the effects of mental model
  soundness on personalizing an intelligent agent}. In
  \bibinfo{booktitle}{\emph{{CHI} '12}} (Austin, USA).
  \bibinfo{publisher}{ACM}, \bibinfo{pages}{1}.
\newblock
\showISBNx{978-1-4503-1015-4}
\urldef\tempurl%
\url{https://doi.org/10.1145/2207676.2207678}
\showDOI{\tempurl}


\bibitem[\protect\citeauthoryear{{Kulesza}, {Stumpf}, {Burnett}, {Yang},
  {Kwan}, and {Wong}}{{Kulesza} et~al\mbox{.}}{2013}]%
        {kulezsa_too_2013}
\bibfield{author}{\bibinfo{person}{Todd {Kulesza}}, \bibinfo{person}{Simone
  {Stumpf}}, \bibinfo{person}{Margaret {Burnett}}, \bibinfo{person}{Sherry
  {Yang}}, \bibinfo{person}{Irwin {Kwan}}, {and} \bibinfo{person}{Weng-Keen
  {Wong}}.} \bibinfo{year}{2013}\natexlab{}.
\newblock \showarticletitle{Too much, too little, or just right? Ways
  explanations impact end users' mental models}. In
  \bibinfo{booktitle}{\emph{VL/HCC '13}}. \bibinfo{pages}{3--10}.
\newblock
\showISSN{1943-6092}
\urldef\tempurl%
\url{https://doi.org/10.1109/VLHCC.2013.6645235}
\showDOI{\tempurl}


\bibitem[\protect\citeauthoryear{Lin, Maire, Belongie, Bourdev, Girshick, Hays,
  Perona, Ramanan, Zitnick, and Dollár}{Lin et~al\mbox{.}}{2014}]%
        {coco}
\bibfield{author}{\bibinfo{person}{Tsung-Yi Lin}, \bibinfo{person}{Michael
  Maire}, \bibinfo{person}{Serge Belongie}, \bibinfo{person}{Lubomir Bourdev},
  \bibinfo{person}{Ross Girshick}, \bibinfo{person}{James Hays},
  \bibinfo{person}{Pietro Perona}, \bibinfo{person}{Deva Ramanan},
  \bibinfo{person}{C.~Lawrence Zitnick}, {and} \bibinfo{person}{Piotr
  Dollár}.} \bibinfo{year}{2014}\natexlab{}.
\newblock \bibinfo{title}{Microsoft COCO: Common Objects in Context}.
\newblock
\newblock
\urldef\tempurl%
\url{http://arxiv.org/abs/1405.0312}
\showURL{%
\tempurl}


\bibitem[\protect\citeauthoryear{Liu, Li, Lee, and Yu}{Liu
  et~al\mbox{.}}{2004}]%
        {liu_04}
\bibfield{author}{\bibinfo{person}{Bing Liu}, \bibinfo{person}{Xiaoli Li},
  \bibinfo{person}{Wee~Sun Lee}, {and} \bibinfo{person}{Philip~S. Yu}.}
  \bibinfo{year}{2004}\natexlab{}.
\newblock \showarticletitle{Text Classification by Labeling Words}. In
  \bibinfo{booktitle}{\emph{AAAI '04}} (San Jose, California).
  \bibinfo{pages}{425–430}.
\newblock


\bibitem[\protect\citeauthoryear{Liu and Avci}{Liu and Avci}{2019}]%
        {liu}
\bibfield{author}{\bibinfo{person}{Frederick Liu} {and} \bibinfo{person}{Besim
  Avci}.} \bibinfo{year}{2019}\natexlab{}.
\newblock \showarticletitle{Incorporating Priors with Feature Attribution on
  Text Classification}. In \bibinfo{booktitle}{\emph{ACL '19}} (Florence,
  Italy). \bibinfo{publisher}{ACL}, \bibinfo{pages}{6274--6283}.
\newblock


\bibitem[\protect\citeauthoryear{Loepp, Barbu, and Ziegler}{Loepp
  et~al\mbox{.}}{2016}]%
        {loepp_interactive_2016}
\bibfield{author}{\bibinfo{person}{Benedikt Loepp},
  \bibinfo{person}{Catalin-Mihai Barbu}, {and} \bibinfo{person}{J\"urgen
  Ziegler}.} \bibinfo{year}{2016}\natexlab{}.
\newblock \showarticletitle{Interactive {Recommending}: {Framework}, {State} of
  {Research} and {Future} {Challenges}}. In
  \bibinfo{booktitle}{\emph{Proceedings of the 1st {Workshop} on {Engineering}
  {Computer}-{Human} {Interaction} in {Recommender} {Systems}}}.
  \bibinfo{pages}{3--13}.
\newblock
\urldef\tempurl%
\url{http://ceur-ws.org/Vol-1705/02-paper.pdf}
\showURL{%
\tempurl}


\bibitem[\protect\citeauthoryear{Loper and Bird}{Loper and Bird}{2002}]%
        {nltk}
\bibfield{author}{\bibinfo{person}{Edward Loper} {and} \bibinfo{person}{Steven
  Bird}.} \bibinfo{year}{2002}\natexlab{}.
\newblock \showarticletitle{NLTK: The Natural Language Toolkit}. In
  \bibinfo{booktitle}{\emph{Proceedings of the ACL Workshop on Effective Tools
  and Methodologies for Teaching Natural Language Processing and Computational
  Linguistics. ACL '02}} (Philadelphia, USA).
\newblock


\bibitem[\protect\citeauthoryear{Lou, Caruana, and Gehrke}{Lou
  et~al\mbox{.}}{2012}]%
        {caruana_2}
\bibfield{author}{\bibinfo{person}{Yin Lou}, \bibinfo{person}{Rich Caruana},
  {and} \bibinfo{person}{Johannes Gehrke}.} \bibinfo{year}{2012}\natexlab{}.
\newblock \showarticletitle{Intelligible Models for Classification and
  Regression}. In \bibinfo{booktitle}{\emph{KDD '12}} (Beijing, China).
  \bibinfo{publisher}{ACM}, \bibinfo{pages}{150–158}.
\newblock
\showISBNx{9781450314626}
\urldef\tempurl%
\url{https://doi.org/10.1145/2339530.2339556}
\showDOI{\tempurl}


\bibitem[\protect\citeauthoryear{Lou, Caruana, Gehrke, and Hooker}{Lou
  et~al\mbox{.}}{2013}]%
        {caruana_1}
\bibfield{author}{\bibinfo{person}{Yin Lou}, \bibinfo{person}{Rich Caruana},
  \bibinfo{person}{Johannes Gehrke}, {and} \bibinfo{person}{Giles Hooker}.}
  \bibinfo{year}{2013}\natexlab{}.
\newblock \showarticletitle{Accurate Intelligible Models with Pairwise
  Interactions}. In \bibinfo{booktitle}{\emph{KDD '13}} (Chicago, USA).
  \bibinfo{publisher}{ACM}, \bibinfo{pages}{623–631}.
\newblock
\showISBNx{9781450321747}
\urldef\tempurl%
\url{https://doi.org/10.1145/2487575.2487579}
\showDOI{\tempurl}


\bibitem[\protect\citeauthoryear{Lundberg and Lee}{Lundberg and Lee}{2017}]%
        {lundberg_2017}
\bibfield{author}{\bibinfo{person}{Scott Lundberg} {and} \bibinfo{person}{Su-In
  Lee}.} \bibinfo{year}{2017}\natexlab{}.
\newblock \showarticletitle{A Unified Approach to Interpreting Model
  Predictions}.
\newblock In \bibinfo{booktitle}{\emph{NIPS '17}}. \bibinfo{publisher}{Curran
  Associates, Inc.}, \bibinfo{pages}{4765--4774}.
\newblock
\urldef\tempurl%
\url{http://papers.nips.cc/paper/7062-a-unified-approach-to-interpreting-model-predictions.pdf}
\showURL{%
\tempurl}


\bibitem[\protect\citeauthoryear{McCarthy}{McCarthy}{1968}]%
        {McCarthy1960ProgramsWC}
\bibfield{author}{\bibinfo{person}{John McCarthy}.}
  \bibinfo{year}{1968}\natexlab{}.
\newblock \bibinfo{booktitle}{\emph{Programs with common sense}}.
\newblock \bibinfo{pages}{403--418}.
\newblock


\bibitem[\protect\citeauthoryear{Mu and Andreas}{Mu and Andreas}{2020}]%
        {mu_2020}
\bibfield{author}{\bibinfo{person}{Jesse Mu} {and} \bibinfo{person}{Jacob
  Andreas}.} \bibinfo{year}{2020}\natexlab{}.
\newblock \showarticletitle{Compositional Explanations of Neurons}. In
  \bibinfo{booktitle}{\emph{Advances in Neural Information Processing
  Systems}}, \bibfield{editor}{\bibinfo{person}{H.~Larochelle},
  \bibinfo{person}{M.~Ranzato}, \bibinfo{person}{R.~Hadsell},
  \bibinfo{person}{M.~F. Balcan}, {and} \bibinfo{person}{H.~Lin}} (Eds.),
  Vol.~\bibinfo{volume}{33}. \bibinfo{publisher}{Curran Associates, Inc.},
  \bibinfo{pages}{17153--17163}.
\newblock
\urldef\tempurl%
\url{https://proceedings.neurips.cc/paper/2020/file/c74956ffb38ba48ed6ce977af6727275-Paper.pdf}
\showURL{%
\tempurl}


\bibitem[\protect\citeauthoryear{O'Donovan, Smyth, Gretarsson, Bostandjiev, and
  H\"{o}llerer}{O'Donovan et~al\mbox{.}}{2008}]%
        {odonovan_2008}
\bibfield{author}{\bibinfo{person}{John O'Donovan}, \bibinfo{person}{Barry
  Smyth}, \bibinfo{person}{Brynjar Gretarsson}, \bibinfo{person}{Svetlin
  Bostandjiev}, {and} \bibinfo{person}{Tobias H\"{o}llerer}.}
  \bibinfo{year}{2008}\natexlab{}.
\newblock \showarticletitle{PeerChooser: Visual Interactive Recommendation}. In
  \bibinfo{booktitle}{\emph{CHI '08}} (Florence, Italy).
  \bibinfo{publisher}{ACM}, \bibinfo{pages}{1085--1088}.
\newblock
\showISBNx{978-1-60558-011-1}
\urldef\tempurl%
\url{https://doi.org/10.1145/1357054.1357222}
\showDOI{\tempurl}


\bibitem[\protect\citeauthoryear{Parra and Brusilovsky}{Parra and
  Brusilovsky}{2015}]%
        {parra_user-controllable_2015}
\bibfield{author}{\bibinfo{person}{Denis Parra} {and} \bibinfo{person}{Peter
  Brusilovsky}.} \bibinfo{year}{2015}\natexlab{}.
\newblock \showarticletitle{User-controllable personalization: {A} case study
  with {SetFusion}}.
\newblock \bibinfo{journal}{\emph{International Journal of Human-Computer
  Studies}}  \bibinfo{volume}{78} (\bibinfo{year}{2015}),
  \bibinfo{pages}{43--67}.
\newblock
\showISSN{10715819}
\urldef\tempurl%
\url{https://doi.org/10.1016/j.ijhcs.2015.01.007}
\showDOI{\tempurl}


\bibitem[\protect\citeauthoryear{Pedregosa, Varoquaux, Gramfort, Michel,
  Thirion, Grisel, Blondel, Prettenhofer, Weiss, Dubourg, Vanderplas, Passos,
  Cournapeau, Brucher, Perrot, and Duchesnay}{Pedregosa et~al\mbox{.}}{2011}]%
        {scikit-learn}
\bibfield{author}{\bibinfo{person}{F. Pedregosa}, \bibinfo{person}{G.
  Varoquaux}, \bibinfo{person}{A. Gramfort}, \bibinfo{person}{V. Michel},
  \bibinfo{person}{B. Thirion}, \bibinfo{person}{O. Grisel},
  \bibinfo{person}{M. Blondel}, \bibinfo{person}{P. Prettenhofer},
  \bibinfo{person}{R. Weiss}, \bibinfo{person}{V. Dubourg}, \bibinfo{person}{J.
  Vanderplas}, \bibinfo{person}{A. Passos}, \bibinfo{person}{D. Cournapeau},
  \bibinfo{person}{M. Brucher}, \bibinfo{person}{M. Perrot}, {and}
  \bibinfo{person}{E. Duchesnay}.} \bibinfo{year}{2011}\natexlab{}.
\newblock \showarticletitle{Scikit-learn: Machine Learning in {P}ython}.
\newblock \bibinfo{journal}{\emph{Journal of Machine Learning Research}}
  \bibinfo{volume}{12} (\bibinfo{year}{2011}), \bibinfo{pages}{2825--2830}.
\newblock


\bibitem[\protect\citeauthoryear{Pu and Chen}{Pu and Chen}{2006}]%
        {pu_2006}
\bibfield{author}{\bibinfo{person}{Pearl Pu} {and} \bibinfo{person}{Li Chen}.}
  \bibinfo{year}{2006}\natexlab{}.
\newblock \showarticletitle{Trust Building with Explanation Interfaces}. In
  \bibinfo{booktitle}{\emph{IUI '06}} (Sydney, AU). \bibinfo{publisher}{ACM},
  \bibinfo{pages}{93--100}.
\newblock
\showISBNx{1-59593-287-9}
\urldef\tempurl%
\url{https://doi.org/10.1145/1111449.1111475}
\showDOI{\tempurl}


\bibitem[\protect\citeauthoryear{Pu, Chen, and Hu}{Pu et~al\mbox{.}}{2011}]%
        {pu_2011}
\bibfield{author}{\bibinfo{person}{Pearl Pu}, \bibinfo{person}{Li Chen}, {and}
  \bibinfo{person}{Rong Hu}.} \bibinfo{year}{2011}\natexlab{}.
\newblock \showarticletitle{A User-centric Evaluation Framework for Recommender
  Systems}. In \bibinfo{booktitle}{\emph{RecSys '11}} (Chicago, USA).
  \bibinfo{publisher}{ACM}, \bibinfo{pages}{157--164}.
\newblock
\showISBNx{978-1-4503-0683-6}
\urldef\tempurl%
\url{https://doi.org/10.1145/2043932.2043962}
\showDOI{\tempurl}


\bibitem[\protect\citeauthoryear{{R Core Team}}{{R Core Team}}{2018}]%
        {rlanguage}
\bibfield{author}{\bibinfo{person}{{R Core Team}}.}
  \bibinfo{year}{2018}\natexlab{}.
\newblock \bibinfo{booktitle}{\emph{R: A Language and Environment for
  Statistical Computing}}.
\newblock R Foundation for Statistical Computing.
\newblock
\urldef\tempurl%
\url{https://www.R-project.org/}
\showURL{%
\tempurl}


\bibitem[\protect\citeauthoryear{Radensky, Downey, Lo, Popovic, and
  Weld}{Radensky et~al\mbox{.}}{2022}]%
        {marissa_22}
\bibfield{author}{\bibinfo{person}{Marissa Radensky}, \bibinfo{person}{Doug
  Downey}, \bibinfo{person}{Kyle Lo}, \bibinfo{person}{Zoran Popovic}, {and}
  \bibinfo{person}{Daniel~S Weld}.} \bibinfo{year}{2022}\natexlab{}.
\newblock \showarticletitle{Exploring the Role of Local and Global Explanations
  in Recommender Systems}. In \bibinfo{booktitle}{\emph{Extended Abstracts of
  the 2022 CHI Conference on Human Factors in Computing Systems}} (New Orleans,
  LA, USA) \emph{(\bibinfo{series}{CHI EA '22})}.
  \bibinfo{publisher}{Association for Computing Machinery},
  \bibinfo{address}{New York, NY, USA}, Article \bibinfo{articleno}{290},
  \bibinfo{numpages}{7}~pages.
\newblock
\showISBNx{9781450391566}
\urldef\tempurl%
\url{https://doi.org/10.1145/3491101.3519795}
\showDOI{\tempurl}


\bibitem[\protect\citeauthoryear{Raghavan and Allan}{Raghavan and
  Allan}{2007}]%
        {raghavan_07}
\bibfield{author}{\bibinfo{person}{Hema Raghavan} {and} \bibinfo{person}{James
  Allan}.} \bibinfo{year}{2007}\natexlab{}.
\newblock \showarticletitle{An Interactive Algorithm for Asking and
  Incorporating Feature Feedback into Support Vector Machines}. In
  \bibinfo{booktitle}{\emph{SIGIR '07}} (Amsterdam, The Netherlands).
  \bibinfo{pages}{79–86}.
\newblock
\showISBNx{9781595935977}
\urldef\tempurl%
\url{https://dl.acm.org/doi/10.1145/1277741.1277758}
\showURL{%
\tempurl}


\bibitem[\protect\citeauthoryear{Ribeiro, Singh, and Guestrin}{Ribeiro
  et~al\mbox{.}}{2016}]%
        {Ribeiro_LIME_2016}
\bibfield{author}{\bibinfo{person}{Marco~Tulio Ribeiro},
  \bibinfo{person}{Sameer Singh}, {and} \bibinfo{person}{Carlos Guestrin}.}
  \bibinfo{year}{2016}\natexlab{}.
\newblock \showarticletitle{"Why Should I Trust You?": Explaining the
  Predictions of Any Classifier}. In \bibinfo{booktitle}{\emph{KDD '16}} (San
  Francisco, USA). \bibinfo{publisher}{ACM}, \bibinfo{pages}{1135--1144}.
\newblock
\showISBNx{978-1-4503-4232-2}
\urldef\tempurl%
\url{https://doi.org/10.1145/2939672.2939778}
\showDOI{\tempurl}


\bibitem[\protect\citeauthoryear{Rieger, Singh, Murdoch, and Yu}{Rieger
  et~al\mbox{.}}{2020}]%
        {rieger}
\bibfield{author}{\bibinfo{person}{Laura Rieger}, \bibinfo{person}{Chandan
  Singh}, \bibinfo{person}{William Murdoch}, {and} \bibinfo{person}{Bin Yu}.}
  \bibinfo{year}{2020}\natexlab{}.
\newblock \showarticletitle{Interpretations are Useful: Penalizing Explanations
  to Align Neural Networks with Prior Knowledge}. In
  \bibinfo{booktitle}{\emph{Proceedings of the 37th International Conference on
  Machine Learning}} \emph{(\bibinfo{series}{Proceedings of Machine Learning
  Research}, Vol.~\bibinfo{volume}{119})},
  \bibfield{editor}{\bibinfo{person}{Hal~Daumé III} {and}
  \bibinfo{person}{Aarti Singh}} (Eds.). \bibinfo{publisher}{PMLR},
  \bibinfo{pages}{8116--8126}.
\newblock
\urldef\tempurl%
\url{http://proceedings.mlr.press/v119/rieger20a.html}
\showURL{%
\tempurl}


\bibitem[\protect\citeauthoryear{Rosenthal and Dey}{Rosenthal and Dey}{2010}]%
        {rosenthal_2010}
\bibfield{author}{\bibinfo{person}{Stephanie Rosenthal} {and}
  \bibinfo{person}{Anind Dey}.} \bibinfo{year}{2010}\natexlab{}.
\newblock \showarticletitle{{Towards Maximizing the Accuracy of Human-Labeled
  Sensor Data}}. In \bibinfo{booktitle}{\emph{IUI '10}}.
  \bibinfo{pages}{259--268}.
\newblock
\showISBNx{9781605585154}


\bibitem[\protect\citeauthoryear{Ross, Hughes, and Doshi-Velez}{Ross
  et~al\mbox{.}}{2017}]%
        {doshivelez}
\bibfield{author}{\bibinfo{person}{Andrew~Slavin Ross},
  \bibinfo{person}{Michael~C. Hughes}, {and} \bibinfo{person}{Finale
  Doshi-Velez}.} \bibinfo{year}{2017}\natexlab{}.
\newblock \showarticletitle{Right for the Right Reasons: Training
  Differentiable Models by Constraining their Explanations}. In
  \bibinfo{booktitle}{\emph{IJCAI '17}}. \bibinfo{pages}{2662--2670}.
\newblock
\urldef\tempurl%
\url{https://doi.org/10.24963/ijcai.2017/371}
\showDOI{\tempurl}


\bibitem[\protect\citeauthoryear{Schaffer, H\"ollerer, and O'Donovan}{Schaffer
  et~al\mbox{.}}{2015}]%
        {schaffer_2015}
\bibfield{author}{\bibinfo{person}{James Schaffer}, \bibinfo{person}{Tobias
  H\"ollerer}, {and} \bibinfo{person}{John O'Donovan}.}
  \bibinfo{year}{2015}\natexlab{}.
\newblock \bibinfo{title}{Hypothetical Recommendation: A Study of Interactive
  Profile Manipulation Behavior for Recommender Systems}.
\newblock
\newblock
\urldef\tempurl%
\url{https://www.aaai.org/ocs/index.php/FLAIRS/FLAIRS15/paper/view/10444}
\showURL{%
\tempurl}


\bibitem[\protect\citeauthoryear{Schapire, Rochery, Rahim, and Gupta}{Schapire
  et~al\mbox{.}}{2005}]%
        {Schapire2005BoostingWP}
\bibfield{author}{\bibinfo{person}{Robert~E. Schapire}, \bibinfo{person}{Marie
  Rochery}, \bibinfo{person}{Mazin~G. Rahim}, {and}
  \bibinfo{person}{Narendra~Kumar Gupta}.} \bibinfo{year}{2005}\natexlab{}.
\newblock \showarticletitle{Boosting with prior knowledge for call
  classification}.
\newblock \bibinfo{journal}{\emph{IEEE Transactions on Speech and Audio
  Processing}}  \bibinfo{volume}{13} (\bibinfo{year}{2005}),
  \bibinfo{pages}{174--181}.
\newblock


\bibitem[\protect\citeauthoryear{Schramowski, Stammer, Teso, Brugger, Shao,
  Luigs, Mahlein, and Kersting}{Schramowski et~al\mbox{.}}{2020}]%
        {schramowski}
\bibfield{author}{\bibinfo{person}{Patrick Schramowski},
  \bibinfo{person}{Wolfgang Stammer}, \bibinfo{person}{Stefano Teso},
  \bibinfo{person}{Anna Brugger}, \bibinfo{person}{Xiaoting Shao},
  \bibinfo{person}{Hans-Georg Luigs}, \bibinfo{person}{Anne-Katrin Mahlein},
  {and} \bibinfo{person}{Kristian Kersting}.} \bibinfo{year}{2020}\natexlab{}.
\newblock \bibinfo{title}{Making deep neural networks right for the right
  scientific reasons by interacting with their explanations}.
\newblock
\newblock
\showeprint[arxiv]{2001.05371}~[cs.LG]


\bibitem[\protect\citeauthoryear{Settles}{Settles}{2009}]%
        {settles2009active}
\bibfield{author}{\bibinfo{person}{Burr Settles}.}
  \bibinfo{year}{2009}\natexlab{}.
\newblock \showarticletitle{Active learning literature survey}.
\newblock  (\bibinfo{year}{2009}).
\newblock


\bibitem[\protect\citeauthoryear{Simard, Amershi, Chickering, Pelton, Ghorashi,
  Meek, Ramos, Suh, Verwey, Wang, and Wernsing}{Simard et~al\mbox{.}}{2017}]%
        {simard_machine_2017}
\bibfield{author}{\bibinfo{person}{Patrice Simard}, \bibinfo{person}{Saleema
  Amershi}, \bibinfo{person}{David Chickering}, \bibinfo{person}{Alicia
  Pelton}, \bibinfo{person}{Soroush Ghorashi}, \bibinfo{person}{Christopher
  Meek}, \bibinfo{person}{Gonzalo Ramos}, \bibinfo{person}{Jina Suh},
  \bibinfo{person}{Johan Verwey}, \bibinfo{person}{Mo Wang}, {and}
  \bibinfo{person}{John Wernsing}.} \bibinfo{year}{2017}\natexlab{}.
\newblock \bibinfo{title}{Machine {Teaching}: {A} {New} {Paradigm} for
  {Building} {Machine} {Learning} {Systems}}.
\newblock
\newblock
\urldef\tempurl%
\url{http://arxiv.org/abs/1707.06742}
\showURL{%
\tempurl}


\bibitem[\protect\citeauthoryear{Sinha, Shen, Song, Ma, Eide, Hsu, and
  Wang}{Sinha et~al\mbox{.}}{2015}]%
        {Sinha_MAG_2015}
\bibfield{author}{\bibinfo{person}{Arnab Sinha}, \bibinfo{person}{Zhihong
  Shen}, \bibinfo{person}{Yang Song}, \bibinfo{person}{Hao Ma},
  \bibinfo{person}{Darrin Eide}, \bibinfo{person}{Bo-June~(Paul) Hsu}, {and}
  \bibinfo{person}{Kuansan Wang}.} \bibinfo{year}{2015}\natexlab{}.
\newblock \showarticletitle{An Overview of Microsoft Academic Service (MAS) and
  Applications}. In \bibinfo{booktitle}{\emph{WWW '15}} (Florence, Italy).
  \bibinfo{publisher}{ACM}, \bibinfo{pages}{243--246}.
\newblock
\showISBNx{978-1-4503-3473-0}
\urldef\tempurl%
\url{https://doi.org/10.1145/2740908.2742839}
\showDOI{\tempurl}


\bibitem[\protect\citeauthoryear{Smith-Renner, Fan, Birchfield, Wu,
  Boyd-Graber, Weld, and Findlater}{Smith-Renner et~al\mbox{.}}{2020}]%
        {SmithRenner2020NoEW}
\bibfield{author}{\bibinfo{person}{Alison Smith-Renner}, \bibinfo{person}{Ron
  Fan}, \bibinfo{person}{M.~Keith Birchfield},
  \bibinfo{person}{Tongshuang~Sherry Wu}, \bibinfo{person}{Jordan~L.
  Boyd-Graber}, \bibinfo{person}{Daniel~S. Weld}, {and} \bibinfo{person}{Leah
  Findlater}.} \bibinfo{year}{2020}\natexlab{}.
\newblock \showarticletitle{No Explainability without Accountability: An
  Empirical Study of Explanations and Feedback in Interactive ML}.
\newblock \bibinfo{journal}{\emph{Proceedings of the 2020 CHI Conference on
  Human Factors in Computing Systems}} (\bibinfo{year}{2020}).
\newblock


\bibitem[\protect\citeauthoryear{Sparck~Jones}{Sparck~Jones}{1988}]%
        {jones}
\bibfield{author}{\bibinfo{person}{Karen Sparck~Jones}.}
  \bibinfo{year}{1988}\natexlab{}.
\newblock \bibinfo{booktitle}{\emph{A Statistical Interpretation of Term
  Specificity and Its Application in Retrieval}}.
\newblock \bibinfo{publisher}{Taylor Graham Publishing},
  \bibinfo{address}{GBR}, \bibinfo{pages}{132–142}.
\newblock
\showISBNx{0947568212}


\bibitem[\protect\citeauthoryear{Sutton and Barto}{Sutton and Barto}{2018}]%
        {sutton_reinforcement_2018}
\bibfield{author}{\bibinfo{person}{Richard~S. Sutton} {and}
  \bibinfo{person}{Andrew~G. Barto}.} \bibinfo{year}{2018}\natexlab{}.
\newblock \bibinfo{booktitle}{\emph{Reinforcement learning: an introduction}
  (\bibinfo{edition}{second edition} ed.)}.
\newblock \bibinfo{publisher}{The MIT Press}, \bibinfo{address}{Cambridge,
  Massachusetts}.
\newblock
\showISBNx{978-0-262-03924-6}


\bibitem[\protect\citeauthoryear{Teso, Alkan, Stammer, and Daly}{Teso
  et~al\mbox{.}}{2022}]%
        {teso_2022_survey}
\bibfield{author}{\bibinfo{person}{Stefano Teso}, \bibinfo{person}{Öznur
  Alkan}, \bibinfo{person}{Wolfang Stammer}, {and} \bibinfo{person}{Elizabeth
  Daly}.} \bibinfo{year}{2022}\natexlab{}.
\newblock \bibinfo{title}{Leveraging Explanations in Interactive Machine
  Learning: An Overview}.
\newblock
\newblock
\urldef\tempurl%
\url{https://doi.org/10.48550/ARXIV.2207.14526}
\showDOI{\tempurl}


\bibitem[\protect\citeauthoryear{{Tintarev} and {Masthoff}}{{Tintarev} and
  {Masthoff}}{2007}]%
        {tintarev_masthoff_07}
\bibfield{author}{\bibinfo{person}{Nava {Tintarev}} {and}
  \bibinfo{person}{Judith {Masthoff}}.} \bibinfo{year}{2007}\natexlab{}.
\newblock \showarticletitle{A Survey of Explanations in Recommender Systems}.
  In \bibinfo{booktitle}{\emph{2007 IEEE 23rd International Conference on Data
  Engineering Workshop}}. \bibinfo{pages}{801--810}.
\newblock


\bibitem[\protect\citeauthoryear{Tintarev and Masthoff}{Tintarev and
  Masthoff}{2011}]%
        {tintarev_designing_2011}
\bibfield{author}{\bibinfo{person}{Nava Tintarev} {and} \bibinfo{person}{Judith
  Masthoff}.} \bibinfo{year}{2011}\natexlab{}.
\newblock \showarticletitle{Designing and {Evaluating} {Explanations} for
  {Recommender} {Systems}}.
\newblock In \bibinfo{booktitle}{\emph{Recommender {Systems} {Handbook}}},
  \bibfield{editor}{\bibinfo{person}{Francesco Ricci}, \bibinfo{person}{Lior
  Rokach}, \bibinfo{person}{Bracha Shapira}, {and} \bibinfo{person}{Paul
  Kantor}} (Eds.). \bibinfo{publisher}{Springer}, \bibinfo{address}{Boston,
  MA}, \bibinfo{pages}{479--510}.
\newblock
\showISBNx{978-0-387-85820-3}
\urldef\tempurl%
\url{https://doi.org/10.1007/978-0-387-85820-3_15}
\showDOI{\tempurl}


\bibitem[\protect\citeauthoryear{Tsai and Brusilovsky}{Tsai and
  Brusilovsky}{2018}]%
        {tsai_2018}
\bibfield{author}{\bibinfo{person}{Chun-Hua Tsai} {and} \bibinfo{person}{Peter
  Brusilovsky}.} \bibinfo{year}{2018}\natexlab{}.
\newblock \showarticletitle{Beyond the Ranked List: User-Driven Exploration and
  Diversification of Social Recommendation}. In \bibinfo{booktitle}{\emph{IUI
  '18}} (Tokyo, Japan). \bibinfo{publisher}{ACM}, \bibinfo{pages}{239--250}.
\newblock
\showISBNx{978-1-4503-4945-1}
\urldef\tempurl%
\url{https://doi.org/10.1145/3172944.3172959}
\showDOI{\tempurl}


\bibitem[\protect\citeauthoryear{Tsai and Brusilovsky}{Tsai and
  Brusilovsky}{2019}]%
        {tsai_social_2019}
\bibfield{author}{\bibinfo{person}{Chun-Hua Tsai} {and} \bibinfo{person}{Peter
  Brusilovsky}.} \bibinfo{year}{2019}\natexlab{}.
\newblock \showarticletitle{Explaining Recommendations in an Interactive Hybrid
  Social Recommender}. In \bibinfo{booktitle}{\emph{IUI '19}} (Marina del Ray,
  USA). \bibinfo{publisher}{ACM}, \bibinfo{pages}{391--396}.
\newblock
\showISBNx{978-1-4503-6272-6}
\urldef\tempurl%
\url{https://doi.org/10.1145/3301275.3302318}
\showDOI{\tempurl}


\bibitem[\protect\citeauthoryear{Tsai and Brusilovsky}{Tsai and
  Brusilovsky}{2020}]%
        {tsai_2020}
\bibfield{author}{\bibinfo{person}{Chun-Hua Tsai} {and} \bibinfo{person}{Peter
  Brusilovsky}.} \bibinfo{year}{2020}\natexlab{}.
\newblock \showarticletitle{The effects of controllability and explainability
  in a social recommender system}.
\newblock \bibinfo{journal}{\emph{User Modeling and User-adapted Interaction}}
  (\bibinfo{year}{2020}), \bibinfo{pages}{1--37}.
\newblock


\bibitem[\protect\citeauthoryear{Verbert, Parra, Brusilovsky, and
  Duval}{Verbert et~al\mbox{.}}{2013}]%
        {verbert_2013}
\bibfield{author}{\bibinfo{person}{Katrien Verbert}, \bibinfo{person}{Denis
  Parra}, \bibinfo{person}{Peter Brusilovsky}, {and} \bibinfo{person}{Erik
  Duval}.} \bibinfo{year}{2013}\natexlab{}.
\newblock \showarticletitle{Visualizing Recommendations to Support Exploration,
  Transparency and Controllability}. In \bibinfo{booktitle}{\emph{IUI '13}}
  (Santa Monica, USA). \bibinfo{publisher}{ACM}, \bibinfo{pages}{351--362}.
\newblock
\showISBNx{978-1-4503-1965-2}
\urldef\tempurl%
\url{https://doi.org/10.1145/2449396.2449442}
\showDOI{\tempurl}


\bibitem[\protect\citeauthoryear{Vig, Sen, and Riedl}{Vig
  et~al\mbox{.}}{2012}]%
        {vig_tag_2012}
\bibfield{author}{\bibinfo{person}{Jesse Vig}, \bibinfo{person}{Shilad Sen},
  {and} \bibinfo{person}{John Riedl}.} \bibinfo{year}{2012}\natexlab{}.
\newblock \showarticletitle{The {Tag} {Genome}: {Encoding} {Community}
  {Knowledge} to {Support} {Novel} {Interaction}}.
\newblock \bibinfo{journal}{\emph{ACM TIIS}} \bibinfo{volume}{2},
  \bibinfo{number}{3} (\bibinfo{year}{2012}), \bibinfo{pages}{1--44}.
\newblock
\showISSN{21606455}
\urldef\tempurl%
\url{https://doi.org/10.1145/2362394.2362395}
\showDOI{\tempurl}


\bibitem[\protect\citeauthoryear{W{\ae}rn}{W{\ae}rn}{2004}]%
        {waern_user_2004}
\bibfield{author}{\bibinfo{person}{Annika W{\ae}rn}.}
  \bibinfo{year}{2004}\natexlab{}.
\newblock \showarticletitle{User {Involvement} in {Automatic} {Filtering}: {An}
  {Experimental} {Study}}.
\newblock \bibinfo{journal}{\emph{UMUAI}} \bibinfo{volume}{14},
  \bibinfo{number}{2} (\bibinfo{year}{2004}), \bibinfo{pages}{201--237}.
\newblock
\showISSN{1573-1391}
\urldef\tempurl%
\url{https://doi.org/10.1023/B:USER.0000028984.13876.9b}
\showDOI{\tempurl}


\bibitem[\protect\citeauthoryear{Wallace, Rodriguez, Feng, Yamada, and
  Boyd-Graber}{Wallace et~al\mbox{.}}{2019}]%
        {wallace_2019_trick}
\bibfield{author}{\bibinfo{person}{Eric Wallace}, \bibinfo{person}{Pedro
  Rodriguez}, \bibinfo{person}{Shi Feng}, \bibinfo{person}{Ikuya Yamada}, {and}
  \bibinfo{person}{Jordan Boyd-Graber}.} \bibinfo{year}{2019}\natexlab{}.
\newblock \showarticletitle{Trick Me If You Can: Human-in-the-Loop Generation
  of Adversarial Examples for Question Answering}.
\newblock \bibinfo{journal}{\emph{Transactions of the Association for
  Computational Linguistics}}  \bibinfo{volume}{7} (\bibinfo{year}{2019}),
  \bibinfo{pages}{387--401}.
\newblock
\urldef\tempurl%
\url{https://doi.org/10.1162/tacl_a_00279}
\showDOI{\tempurl}


\bibitem[\protect\citeauthoryear{Wang, Kale, Nori, Stella, Nunnally, Chau,
  Vorvoreanu, Vaughan, and Caruana}{Wang et~al\mbox{.}}{2021}]%
        {wang2021}
\bibfield{author}{\bibinfo{person}{Zijie~Jay Wang}, \bibinfo{person}{Alex
  Kale}, \bibinfo{person}{Harsha Nori}, \bibinfo{person}{Peter Stella},
  \bibinfo{person}{Mark~E. Nunnally}, \bibinfo{person}{Duen~Horng Chau},
  \bibinfo{person}{Mihaela Vorvoreanu}, \bibinfo{person}{Jennifer~Wortman
  Vaughan}, {and} \bibinfo{person}{Rich Caruana}.}
  \bibinfo{year}{2021}\natexlab{}.
\newblock \showarticletitle{GAM Changer: Editing Generalized Additive Models
  with Interactive Visualization}.
\newblock \bibinfo{journal}{\emph{ArXiv}}  \bibinfo{volume}{abs/2112.03245}
  (\bibinfo{year}{2021}).
\newblock


\bibitem[\protect\citeauthoryear{Weld and Bansal}{Weld and Bansal}{2019}]%
        {weld_challenge_2018}
\bibfield{author}{\bibinfo{person}{Daniel Weld} {and} \bibinfo{person}{Gagan
  Bansal}.} \bibinfo{year}{2019}\natexlab{}.
\newblock \showarticletitle{The Challenge of Crafting Intelligible
  Intelligence}.
\newblock \bibinfo{journal}{\emph{CACM}} \bibinfo{volume}{62},
  \bibinfo{number}{6} (\bibinfo{year}{2019}), \bibinfo{pages}{70--79}.
\newblock
\showISSN{0001-0782}
\urldef\tempurl%
\url{https://doi.org/10.1145/3282486}
\showDOI{\tempurl}


\bibitem[\protect\citeauthoryear{Wu, Weld, and Heer}{Wu et~al\mbox{.}}{2019}]%
        {sherry_tochi}
\bibfield{author}{\bibinfo{person}{Tongshuang Wu}, \bibinfo{person}{Daniel~S.
  Weld}, {and} \bibinfo{person}{Jeffrey Heer}.}
  \bibinfo{year}{2019}\natexlab{}.
\newblock \showarticletitle{Local Decision Pitfalls in Interactive Machine
  Learning: An Investigation into Feature Selection in Sentiment Analysis}.
\newblock \bibinfo{journal}{\emph{ACM Trans. Comput.-Hum. Interact.}}
  \bibinfo{volume}{26}, \bibinfo{number}{4}, Article \bibinfo{articleno}{24}
  (\bibinfo{date}{jun} \bibinfo{year}{2019}), \bibinfo{numpages}{27}~pages.
\newblock
\showISSN{1073-0516}
\urldef\tempurl%
\url{https://doi.org/10.1145/3319616}
\showDOI{\tempurl}


\bibitem[\protect\citeauthoryear{Wu and Srihari}{Wu and Srihari}{2004}]%
        {wu_04}
\bibfield{author}{\bibinfo{person}{Xiaoyun Wu} {and} \bibinfo{person}{Rohini
  Srihari}.} \bibinfo{year}{2004}\natexlab{}.
\newblock \showarticletitle{Incorporating Prior Knowledge with Weighted Margin
  Support Vector Machines}. In \bibinfo{booktitle}{\emph{KDD '04}} (Seattle,
  USA). \bibinfo{pages}{326–333}.
\newblock
\showISBNx{1581138881}
\urldef\tempurl%
\url{https://doi.org/10.1145/1014052.1014089}
\showDOI{\tempurl}


\bibitem[\protect\citeauthoryear{Zhang and Chen}{Zhang and Chen}{2018}]%
        {zhang_explainable_2018}
\bibfield{author}{\bibinfo{person}{Yongfeng Zhang} {and} \bibinfo{person}{Xu
  Chen}.} \bibinfo{year}{2018}\natexlab{}.
\newblock \showarticletitle{Explainable {Recommendation}: {A} {Survey} and
  {New} {Perspectives}}.
\newblock \bibinfo{journal}{\emph{Arxiv}} (\bibinfo{year}{2018}).
\newblock
\urldef\tempurl%
\url{http://arxiv.org/abs/1804.11192}
\showURL{%
\tempurl}


\end{thebibliography}

\end{document}